\documentclass[8pt,nofootinbib,showpacs,superscriptaddress,titlepage,floatfix,amsmath,amssymb,preprint,longbibliography]{revtex4-1}
\usepackage{dcolumn}
\usepackage{subfigure}
\usepackage{bm}
\usepackage{amssymb}
\usepackage{amsmath}
\usepackage{colordvi}
\usepackage{graphicx}
\usepackage{color}
\usepackage[colorlinks=true,linkcolor=blue,citecolor=blue,urlcolor=blue]{hyperref}%
\usepackage{hyperref}
\usepackage{comment}
\usepackage{float}
\usepackage{harpoon}
\usepackage{soul}
\usepackage[T1]{fontenc}
\UseRawInputEncoding
\newcommand{\be}{\begin{equation}}
\newcommand{\ee}{\end{equation}}
\begin{document}

\title{Moderate-temperature near-field thermophotovoltaic systems with thin-film InSb cells}

\author{Rongqian Wang}\email{rongqianwang@sina.com}
\address{School of physical science and technology \& Collaborative Innovation Center of Suzhou Nano Science and Technology, Soochow University, Suzhou 215006, China.}

\author{Jincheng Lu}
\affiliation{School of physical science and technology \&
	Collaborative Innovation Center of Suzhou Nano Science and Technology, Soochow University, Suzhou 215006, China.}
\address{Center for Phononics and Thermal Energy Science, China-EU Joint Center for Nanophononics, Shanghai Key Laboratory of Special Artificial Microstructure Materials and Technology, School of Physics Science and Engineering, Tongji University, Shanghai 200092 China}

\author{Jian-Hua Jiang}\email{jianhuajiang@suda.edu.cn}
\address{School of physical science and technology \& Collaborative Innovation Center of Suzhou Nano Science and Technology, Soochow University, Suzhou 215006, China.}


\date{\today}
\begin{abstract}
Near-field thermophotovoltaic systems functioning at 400$\sim$900~K based on graphene-hexagonal-boron-nitride heterostructures and thin-film InSb $p$-$n$ junctions are investigated theoretically. The performances of two near-field systems with different emitters are examined carefully. One near-field system consists of a graphene-hexagonal-boron-nitride-graphene sandwich structure as the emitter, while the other system has an emitter made of the double graphene-hexagonal-boron-nitride heterostructure. It is shown that both systems exhibit higher output power density and energy efficiency than the near-field system based on mono graphene-hexagonal-boron-nitride heterostructure. The optimal output power density of the former device can reach to $1.3\times10^{5}~\rm{W\cdot m^{-2}}$, while the optimal energy efficiency can be as large as $42\%$ of the Carnot efficiency. We analyze the underlying physical mechanisms that lead to the excellent performances of the proposed near-field thermophotovoltaic systems. Our results are valuable toward high-performance moderate temperature thermophotovoltaic systems as appealing thermal-to-electric energy conversion (waste heat harvesting) devices.
\end{abstract}

\maketitle


\section{Introduction}
Thermophotovoltaic (TPV) systems are solid-state renewable energy resource that are of immense potentials in a wide range of applications including solar energy harvesting and waste heat recovery~\cite{shockley1961detailed,martin2004temperature,nagashima2007germanium,fraas2000three,sulima2001fabrication,wu2012metamaterial,chan2013toward,liao2016efficiently,zhao2017high,Tervo2018}.
In the TPV system, a photovoltaic (PV) cell is placed in the proximity of a thermal emitter and converts the thermal radiation from the emitter into electricity via infrared photoelectric conversion. However, the frequency mismatch between a moderate-temperature thermal emitter and the PV cell leads to significantly reduced performance of the TPV systems at moderate temperatures (i.e., 400$\sim$900~K which is the majority spectrum of the industry waste heat). To overcome this obstacle, materials which support surface polaritons have been used to introduce a resonant near-field energy exchange between the emitter and the absorber~\cite{wu2012metamaterial,svetovoy2012plasmon,
ilic2012overcoming,svetovoy2014graphene,basu2015near,zhao2017high}. As a consequence, near-field TPV (NTPV) systems have been proposed to achieve appealing energy efficiency and output power~\cite{narayanaswamy2003surface,laroche2006JAP,park2008performance,ilic2012overcoming,bright2014performance,molesky2015ideal,st2017hot,Jiang2018Near,papadakis2020broadening}. 
Near-field systems based on graphene, hexagonal-boron-nitride ({\it h}-BN) and their heterostructures have been shown to demonstrate excellent near-field couplings due to surface plasmon polaritons (SPPs), surface phonon polaritons (SPhPs) and their hybridizations [i.e., surface plasmon-phonon polaritons (SPPPs)]~\cite{svetovoy2012plasmon,messina2013graphene,svetovoy2014graphene,woessner2015retime,Bo_JHT,Bo_PRB,Sailing_ACS}. It was demonstrated that a heterostructure consisting of graphene-{\it h}-BN multilayers performs better than the monocell structure and the heat flux is found to be three times larger than that of the monocell structure~\cite{Bo_JHT,Bo_PRB,Sailing_ACS}. Here, we consider the graphene-{\it h}-BN-graphene sandwich structure and graphene-{\it h}-BN-graphene-{\it h}-BN four-layer structure as the near-field thermal emitters to provide the enhanced radiative heat transfer.

In order to convert the infrared radiation into electricity, the energy bandgap ($E_{\rm gap}$) of the NTPV cell (i.e., the $p$-$n$ junction) must match the radiative spectrum generated by the emitter. III-V group compound semiconductors like Gallium Arsenide (GaAs), Gallium antimonide(GaSb), Indium antimonide (InSb) and Indium Arsenide (InAs) have been used due to the small bandgap energy, high electron mobility and low electron effective mass~\cite{ilic2012overcoming,laroche2006JAP,messina2013graphene,zhao2017high}.
Recently, semiconductor thin-films have been explored in NTPV systems. In Refs.~\cite{zhao2017high} and \cite{papadakis2020broadening}, a NTPV system based on a InAs thin-film cell has exhibits appealing performance operating at high temperatures. But the system suffers from low energy efficiency (below $10\%$) when operating at moderate temperature due to the parasitic heat transfer induced by the phonon-polaritons of InAs. Here, we use InSb as the near-field absorber since the bandgap energy of InSb is lower compared to InAs and its photon-phonon interaction is much weaker than InAs. For the temperature of the InSb cell, $T_{\rm cell}=320~{\rm K}$, which has been proved an optimal cell temperature in our previous work~\cite{PRAppliedWang}, the gap energy is $E_{\rm gap}=0.17$~eV and the corresponding angular frequency is $\omega_{\rm gap} = 2.5\times10^{14}$~rad/s.

In this work, we examine the performances of two NTPV devices: the graphene-$h$-BN-graphene-InSb cell (denoted as G-FBN-G-InSb cell, with the graphene-$h$-BN-graphene sandwich structure being the emitter and the InSb thin-film being the cell) and the graphene-$h$-BN-graphene-$h$-BN-InSb cell (denoted as G-FBN-G-FBN-InSb cell, with the double graphene-$h$-BN heterostructure being the emitter and the InSb thin-film being the cell). We study and compare the performances of these two systems and reveal their underlying physical mechanisms. We further optimize the performance of the near-field TPV systems for various parameters.

In this work, we address two issues: First, we try to optimize the design of the {\it h}-BN-graphene heterostructures as the emitter to improve the performance of the NTPV system. Second, we try to discuss the effect of finite thickness of the InSb cell on the performance of the NTPV system.

This work is structured as follows. In Sec.~\ref{System and Theory}, we describe our NTPV system and clarify the radiative heat flux exchanged between the emitter and the cell. In Sec.~\ref{nfperformance}, we
study and compare the performances of the two NTPV systems and analyze the spectral distributions of the photo-induced current density and indicent heat flux. We also study the photon tunneling coefficient to further elucidate the physical mechanisms. In Sec.~\ref{nfoptimization}, we examine the performances of the two NTPV systems for various InSb thin-film thicknesses and emitter temperatures to optimize the performances of the NTPV systems. Finally, we summarize and conclude in Sec.~\ref{conclusions}.

\begin{figure}
\centering\includegraphics[width=12cm]{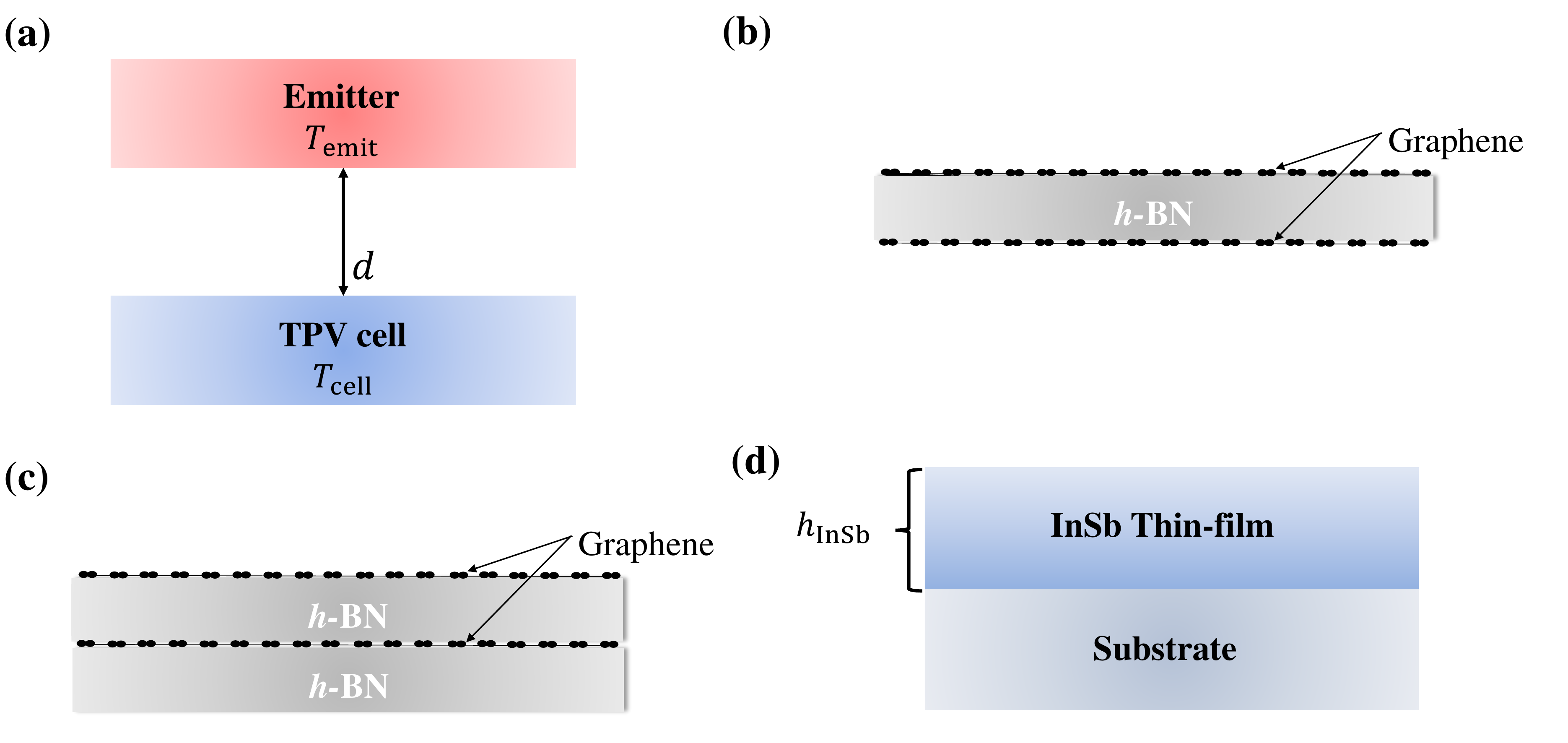}
\caption{Schematic illustration of the NTPV systems. (a) A representative TPV device. A thermal emitter of temperature $T_{\rm emit}$ is located at a distance $d$ of a thermophotovoltaic cell of temperature $T_{\rm cell}$. Two compositions of the thermal emitter with (b) graphene-{\it h}-BN-graphene structure and (c) graphene-{\it h}-BN-graphene-$h$-BN structure. (d) The thermophotovoltaic cell with a InSb thin-film of thickness $h_{\rm InSb}$ supported by a semi-infinite substrate.}
~\label{model}
\end{figure}

\section{System and Theory}\label{System and Theory}
In Fig.~\ref{model}, we consider the graphene-{\it h}-BN-InSb NTPV systems. Fig.~\ref{model}(a) is a schematic presentation of a typical NTPV system, which consists of a thermal emitter and a thermophotovoltaic cell. The emitter and the thermophotovoltaic cell is separated by a vacuum gap with thickness $d$. The temperatures of the emitter and cell are kept at $T_{\rm emit}$ and $T_{\rm cell}$, respectively. The thermal radiation from the emitter is absorbed by the cell and then converted into electricity via photoelectric conversion.
Figs.~\ref{model}(b) and \ref{model}(c) present the two compositions of the emitter. Fig.~\ref{model}(b) is a graphene-{\it h}-BN-graphene sandwich structure and Fig.~\ref{model}(c) is made of two graphene-{\it h}-BN heterostructures. The thickness of {\it h}-BN thin film is $h_{\rm BN}$ and the graphene monolayer is model as a layer of thickness $h_{\rm g}$.
Fig.~\ref{model}(d) is a thin-film InSb $p$-$n$ junction supported by a substrate. The thickness of the InSb thin-film is $h_{\rm InSb}$ and the substrate is set to be semi-infinite intrinsic InSb.

When the InSb cell is located at a distance $d$ which is on the order of or smaller than the thermal wavelength $\lambda_{\rm th}=2\pi \hbar c/k_{\rm B}T_{\rm emit}$ from the emitter, the thermal radiation can be significantly enhanced due to energy transfer via evanescent waves~\cite{ilic2012overcoming}. The radiative heat exchange between the emitter and the cell is given by~\cite{polder1971theory,pendry1999radiative}
\begin{align}
 Q_{\rm rad}= Q_{\omega<\omega_{\rm gap}} + Q_{\omega\ge\omega_{\rm gap}} \label{Qrad}
\end{align}
where $Q_{\omega<\omega_{\rm gap}}$ and $Q_{\omega\ge\omega_{\rm gap}}$ are the heat exchanges below and above the band gap of the cell, respectively.

The below  $Q_{\omega<\omega_{\rm gap}}$ and above-gap heat exchange $Q_{\omega\ge\omega_{\rm gap}}$ are respectively given by
\begin{align}
& Q_{\omega<\omega_{\rm gap}}=\int_{0}^{\omega_{\rm gap}}\frac{d\omega}{4\pi^2}\left[\Theta_{1}\left(T_{\rm emit},\omega\right)-\Theta_{2}\left(T_{\rm cell},\omega\right)\right]\sum_{j}\int kdk \zeta_j(\omega,k), \label{Qblow}
\end{align}
and
\begin{align}
& Q_{\omega\ge\omega_{\rm gap}}=\int_{\omega_{\rm gap}}^{\infty}\frac{d\omega}{4\pi^2}\left[\Theta_{1}\left(T_{\rm emit},\omega \right)-\Theta_{2}\left(T_{\rm cell},\omega,\Delta\mu \right)\right]\sum_{j}\int kdk \zeta_j(\omega,k), \label{Qabove}
\end{align}
where $\Theta_1\left(T_{\rm emit},\omega\right)={\hbar\omega}/[\exp{\left(\frac{\hbar\omega}{k_{\rm B}T_{\rm emit}}\right)}-1]$ and $\Theta_2\left(T_{\rm cell},\omega,\Delta\mu\right)={\hbar\omega}/{[\exp{\left(\frac{\hbar\omega-\Delta\mu}{k_{\rm B}T_{\rm cell}}\right)}-1]}$ are the Planck mean oscillator energies of blackbody at temperature $T_{\rm emit}$ and $T_{\rm cell}$, respectively. $\Delta\mu$ is the electrochemical potential difference across the $p$-$n$ junction, which describes the effects of charge injection or depletion on the carrier recombination processes and hence modify the number of photons through the detailed balance. $k$ is the magnitude of
the in-plane wavevector of thermal radiation waves.
$\zeta_j(\omega,k)$ is the photon transmission coefficient for the $j$-th polarization $(j=s,p)$, which consists of the contributions from both the propagating and the evanescent waves~\cite{mulet2002enhanced}
\begin{equation}
\zeta_j(\omega,k)=
\begin{cases}
\frac{\left(1-\left|r_{\rm emit}\right|^2\right)\left(1-\left|r_{\rm cell}\right|^2\right)}{\left|1-r_{\rm emit}^j r_{\rm cell}^j \exp(2ik_z^0d)\right|^2}, & k<\frac{\omega}{c} \\
             \frac{4{\rm Im}\left(r_{\rm emit}^j\right){\rm Im}\left(r_{\rm cell}^j\right)\exp(2ik_z^0d)}{\left|1-r_{\rm emit}^j r_{\rm cell}^j \exp(2ik_z^0d)\right|^2}, & k\ge\frac{\omega}{c} \label{photon tunneling}
\end{cases}
\end{equation}
where $k_{z}^{0}=\sqrt{\omega^2/c^2-k^2}$ is the perpendicular-to-plane component of the wavevector in vacuum. $r_{\rm emit}^j$ ($r_{\rm cell}^j$) with $j=s,p$ is the complex reflection coefficient at the interface between the emitter (cell) and the air. Here, the reflection coefficients of the emitter and cell are calculated by the scattering matrix approach~\cite{scattering1999,Zhang2007Nano}.

Via the infrared photoelectric conversion, the above-gap  radiative heat exchange is then converted into electricity. Based on the detailed balance analysis, the electric current density of a NTPV cell is given by~\cite{shockley1961detailed,ashcroft2010IVcurve}
\begin{equation}
\begin{aligned}
I=I_{\rm ph}-I_0[\exp(V/V_{\rm cell})-1],  \label{Ie}
\end{aligned}
\end{equation}
where $V=\Delta\mu/e$ is the voltage bias across the NTPV cell, $V_{\rm cell}=k_{\rm B}T_{\rm cell}/e$ is a voltage which measures the temperature of the cell~\cite{shockley1961detailed}. $I_{\rm ph}$ and $I_0$ are the photo-generation current density and reverse saturation current density, respectively. In Eq.~(\ref{Ie}), an ideal rectifier condition has been used to  simplify the nonradiative recombination~\cite{shockley1961detailed}. The actual nonradiative mechanisms include Shockley-Read-Hall (RSH) and Auger nonradiative processes. Here, for the sake of simplicity, we just follow the Shockley-Queisser analysis~\cite{shockley1961detailed}.

The reverse saturation current density is determined by the diffusion of minority carriers in the InSb $p$-$n$ junction, which is given by
\begin{equation}
\begin{aligned}
I_0=en_{\rm i}^2\left(\frac{1}{N_{\rm A}}\sqrt{\frac{D_{\rm e}}{\tau_{\rm e}}}+\frac{1}{N_{\rm D}}\sqrt{\frac{D_{\rm h}}{\tau_{\rm h}}} \right),
\end{aligned}
\end{equation}
where $n_{\rm i}$ is the intrinsic carrier concentration, $D_{\rm e}$ and $D_{\rm h}$ are the diffusion coefficients of the electrons and holes, respectively. $N_{\rm A}$ and $N_{\rm D}$ are the $p$-region and $n$-region impurity concentrations, respectively~\cite{shur1996handbook}. $\tau_{\rm e}$ and $\tau_{h}$ are correspondingly the relaxation time of the electron-hole pairs in the $n$-region and $p$-region. Numerical values of these parameters are taken from Refs.~\cite{shur1996handbook} and \cite{lim2015graphene}.

The  photo-generation current density is contributed from the above-gap thermal heat exchange~\cite{laroche2006JAP,messina2013graphene}
\begin{equation}
\begin{aligned}
& I_{\rm ph}=\frac{e}{4\pi^2}\int_{\omega_{\rm gap}}^{\infty}\frac{d\omega}{\hbar\omega}\left[\Theta_{1}\left(T_{\rm emit},\omega \right)-\Theta_{2}\left(T_{\rm cell},\omega,\Delta\mu \right)\right]\\
& \hspace{1 cm} \times\sum_{j}\int kdk \zeta_j(\omega,k)\left(1-\exp{\left[-2\rm Imag(\it k_{z}^{\rm InSb})\it h_{\rm InSb}\right]}\right), \label{Iph}
\end{aligned}
\end{equation}
where $\it k_{z}^{\rm InSb}=\sqrt{\varepsilon_{\rm InSb}\omega^2/c^2-k^2}$ is the perpendicular-to-plane component of the wavevector in the InSb $p$-$n$ junction. $\varepsilon_{\rm InSb}$ is the dielectric function of the InSb cell, which is given by $\varepsilon_{\rm InSb} = \left(n+\frac{i\rm c \it\alpha(\omega)}{2\omega}\right)^2$. $n_{\rm InSb}=4.12$ is the refractive index and $c$ is the speed of light in vacuum.
$\alpha(\omega)$ is a step-like function
describing the photonic absorption, which is given by~\cite{messina2013graphene} $\alpha(\omega)=0$ for
$\omega < \omega_{\rm gap}$ and $\alpha(\omega)=\alpha_0\sqrt{\omega/\omega_{\rm gap}-1}$ for
$\omega > \omega_{\rm gap}$ with $\alpha_0$ = 0.7 $\mu\rm m^{-1}$~\cite{messina2013graphene}.
Since the NTPV cell is a thin film,  the exponential decay characteristic of the electromagnetic wave propagating in the InSb thin-film must be considered. The term $\left(1-\exp{\left[-2\rm Imag(\it k_{z}^{\rm InSb})\it h_{\rm InSb}\right]}\right)$ is the absorption probability of the incident radiation in the InSb film of thickness $h_{\rm InSb}$, which measures the actual availability of the above-gap photons in the photon-carrier generation process.

The dielectric function of {\it h}-BN is described by a Drude-Lorentz
model, which is given by~\cite{SPPPs2}
\begin{equation}
\varepsilon_m = \varepsilon_{\infty,m}\left(1+ \frac{\omega^2_{\rm LO,\it m}-\omega^2_{\rm TO,\it m}}{\omega^2_{\rm TO,\it m}-i\gamma_m\omega-\omega^2}\right),
\end{equation}
where $m = {\rm \parallel, \perp}$ denotes the out-of-plane and the in-plane directions, respectively.  $\varepsilon_{\infty,m}$ is the high-frequency relative permittivity, $\omega_{\rm TO}$ and $\omega_{\rm LO}$ are the transverse and longitudinal optical phonon frequencies, respectively. $\gamma_m$ is the damping constant of the optical phonon modes. The values of these parameters are chosen as those determined by experiments~\cite{geick1966normal,expara_hBN}.

The effective dielectric function of the graphene monolayer is modeled as~\cite{Egraphene}
\begin{equation}
\varepsilon_{\rm g} = 1+i\frac{\sigma_{\rm g}}{\varepsilon_{0}\omega h_{\rm g}},
\end{equation}
where $\sigma_{\rm g}$ is the optical conductivity~\cite{Cgraph}.

The output electric power density $P_{\rm e}$ of the NTPV system is defined as the product of the net electric current density and the voltage bias,
\begin{equation}
\begin{aligned}
P_{\rm e}= -I_{\rm e}V, \label{epower}
\end{aligned}
\end{equation}
and the energy efficiency $\eta$ is given by the ratio between the output electric power density $P_{\rm e}$ and incident radiative heat flux $Q_{\rm inc}$,
\begin{equation}
\begin{aligned}
\eta= \frac{P_{\rm e}}{Q_{\rm inc}},\label{effi}
\end{aligned}
\end{equation}
where the incident radiative heat flux is given by the radiative heat exchange defined in Eq.~\eqref{Qrad}.
The second law of thermodynamics imposes an upper bound on the energy efficiency, which is the Carnot efficiency,
\begin{equation}
\begin{aligned}
\eta_c= 1- \frac{T_{\rm cell}}{T_{\rm emit}}. 
\end{aligned}
\end{equation}
When studying the energy efficiency of various NTPV systems with different temperatures, we use the Carnot efficiency to quantify and compare their energy efficiencies.

\section{Performances of graphene-{\it h}-BN-InSb near-field systems}\label{nfperformance}
\begin{figure}
\centering\includegraphics[width=12cm]{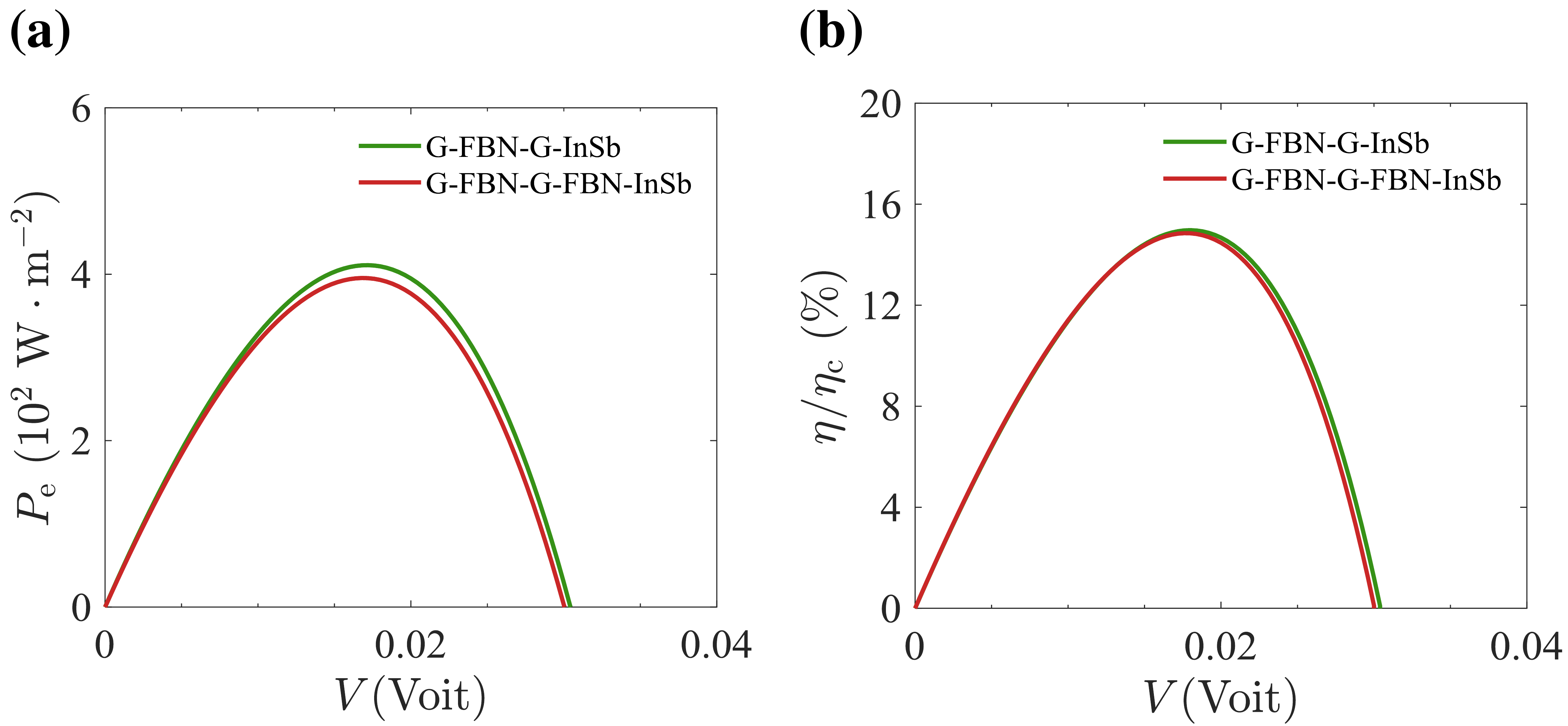}
\caption{Performances of two NTPV cells. (a) Electrical power density and (b) energy$-$conversion efficiency in units of Carnot efficiency ($\eta_{\rm c}$). The temperatures of the emitter and the cell are set at $T_{\rm emit}=450$ K and $T_{\rm cell}=320$ K, respectively.
The vacuum gap distance is $d=20$ nm, the {\it h}-BN film thickness is $h_{\rm BN}=20$ nm and the thickness of the InSb thin film is $h_{\rm InSb}=1000$ nm. The chemical potential of graphene is $\mu_{\rm g}=1.0$ eV. The Carnot efficiency is given by $\eta_{\rm c}=1-{T_{\rm cell}}/{T_{\rm emit}}$.}\label{fig:performance}
\end{figure}

We first examine the performances of the two types of NTPV cells. Fig.~\ref{fig:performance} shows the output power density and energy efficiency as a function of voltage bias for the two set-ups, respectively denoted as G-FBN-G-InSb cell (graphene-{\it h}-BN-graphene  heterostructure as the emitter and thin-film InSb $p$-$n$ junction as the receiver) and G-FBN-G-FBN-InSb cell (graphene-{\it h}-BN-graphene-{\it h}-BN heterostructure as the emitter and thin-film InSb $p$-$n$ junction as the receiver). The output power density and energy efficiency are optimized for various physical parameters, including the temperature of the cell, the chemical potential of graphene and {\it h}-BN film thickness~\cite{PRAppliedWang}. The analysis in Ref.~[\onlinecite{PRAppliedWang}] shows that setting $\mu_{\rm g}=1.0$~eV, $h_{\rm BN}=20$~nm and $T_{\rm cell}=320$~K provides roughly optimal performance, both in terms of power and efficiency, for the NTPV systems considered in this work. Therefore, these parameters are kept as those constants in the main text. The thickness of the InSb thin film is set as $h_{\rm InSb}=1000$ nm, which is close to the thickness of experimental samples~\cite{yang2006optical}.

As shown in Fig.~\ref{fig:performance}, the maximum output power density of the G-FBN-G-InSb cell is about $4.1\times10^{2}~\rm{W\cdot m^{-2}}$, nearly 1.1 times of the the maximum output power density of the G-FBN-G-FBN-InSb cell (about $3.9\times10^{2}~\rm{W\cdot m^{-2}}$). For the energy-conversion efficiency, the maximum values of the G-FBN-G-InSb cell is about $15\%\eta_c$, which slightly higher than the maximum efficiency of the FBN-G-FBN-InSb cell (about $14.9\%\eta_c$).

\begin{figure}
\centering\includegraphics[width=12cm]{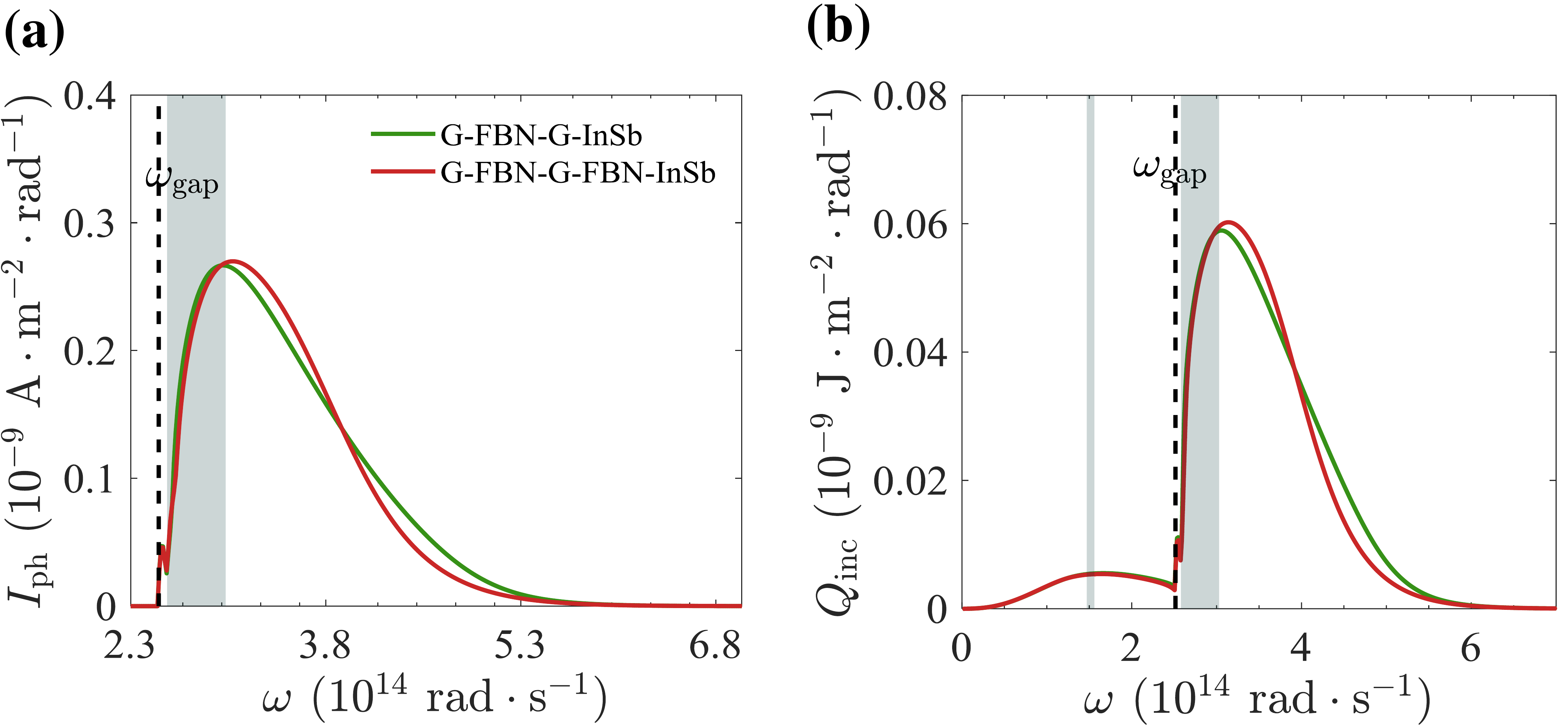}
\caption{(a) Photo-induced current spectra $I_{\rm ph}(\omega)$ and (b) incident radiative heat spectra $Q_{\rm inc}(\omega)$ at the maximum electric power density for the two configurations. The temperatures of the emitter and the cell are kept at $T_{\rm emit}=450$~K and $T_{\rm cell}=320$~K, respectively. The vacuum gap is $d=20$ nm and the chemical potential of graphene is set as $\mu_{\rm g}=1.0$~eV. The {\it h}-BN film thickness is $h_{\rm film}=20$~nm. The chemical potential difference across the InSb $p$-$n$ junction $\Delta \mu$ is optimized independently for maximum output power for each configuration.}\label{fig:spectrum}
\end{figure}

In order to analyze the physical mechanisms responsible for such performances of these two setups, we show the spectral distributions of the photo-induced current and incident radiative heat flux at the maximum electric power density for the G-FBN-G-InSb cell and G-FBN-G-FBN-InSb cell in Fig.~\ref{fig:spectrum}. The two shaded areas in Fig.~\ref{fig:spectrum}(a) and (b) are the two reststrahlen bands of {\it h}-BN~\cite{jacob2014nanophotonics}. As exhibited in Fig.~\ref{fig:spectrum}(a) that higher photo-induced current spectra of the two systems appear at and near the reststrahlen band due to the reststrahlen effect. 
The reststrahlen effect originates from the hyperbolic optical properties of the {\it h}-BN film due to the photon---optical-phonon interactions. In the reststrahlen band presented in Fig.~\ref{fig:spectrum}(a), the in-plane dielectric function of {\it h}-BN is negative, leading to strong reflection and suppressed transmission of incident photons~\cite{jacob2014nanophotonics}. The near-field radiation effects are essentially caused by the evanescent propagation of the incident photons, which dramatically enhances the radiative heat transfer between two closed spaced objects~\cite{PRAppliedWang,SPPPs1,SPPPs2,woessner2015retime,Bo_JHT,Bo_PRB,Sailing_ACS}. Such enhancement of the radiative heat transfer eventually leads to the significant increase of the input heat flux, output electric power and the energy efficiency~\cite{PRAppliedWang}.

It is noticed that the overall photo-induced current spectrum of the G-FBN-G-InSb cell is higher than the one of the G-FBN-G-FBN-InSb cell except in the frequency range from $3.1\times10^{14}~\rm{rad \cdot s^{-1}}$ to $4.1\times10^{14}~\rm{rad\cdot s^{-1}}$. After integrating over $\omega$, this higher photo-induced current spectrum gives rise to improved output power of the G-FBN-G-InSb cell.

Contrary to the photo-induced current spectra, the overall incident radiative heat flux of the G-FBN-G-FBN-InSb cell is higher than the G-FBN-G-InSb cell. Since the energy-conversion efficiency is defined as the ratio of the output electric power density and the incident radiation heat flux, the higher output power density and lower incident radiation heat flux give a higher energy efficiency of the G-FBN-G-InSb cell.

\begin{figure}
\centering\includegraphics[width=12cm]{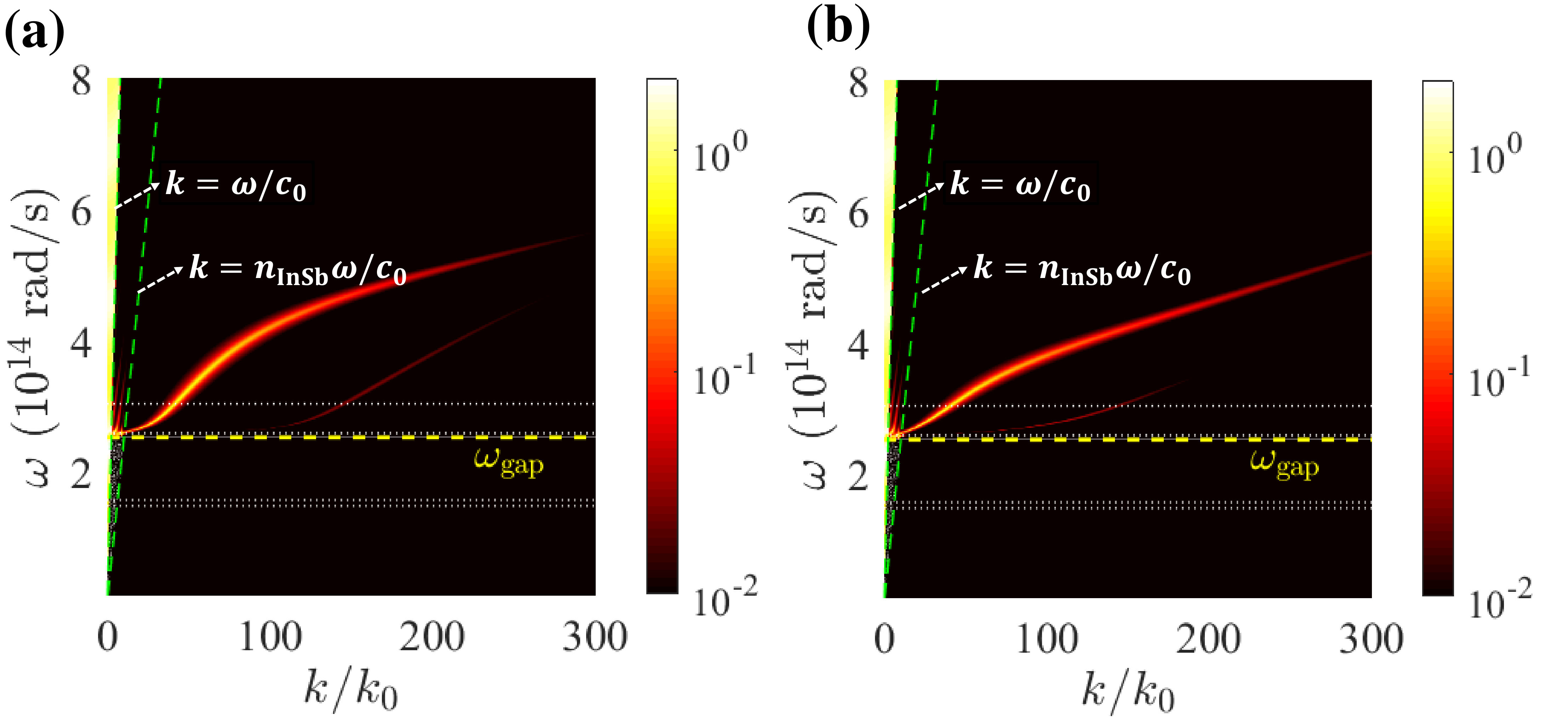}
\caption{Photon transmission coefficient $\zeta(\omega,k)$ for (a) G-FBN-G-InSb cell and (b) G-FBN-G-FBN-InSb cell.  The temperatures of the emitter and the cell are kept at $T_{\rm emit}=450$~K and $T_{\rm cell}=320$~K, respectively. The vacuum gap is $d=20$ nm and the chemical potential of graphene is set as $\mu_{\rm g}=1.0$~eV. The {\it h}-BN film thickness is $h_{\rm film}=20$~nm and the InSb thin-film thickness is $h_{\rm InSb}=1000$ nm.}\label{fig:photon tunneling}
\end{figure}

To further elucidate the mechanism for the enhanced performance of the near-field systems, we examine the photon tunneling coefficient $\zeta_j(\omega,k)$ (given by Eq.~\ref{photon tunneling}). As shown in Fig. ~\ref{fig:photon tunneling}, the bright color indicates a high transmission coefficient. The green dashed lines are the light lines of vacuum and InSb, respectively. Note that the maximum transmission coefficient is 2 due to the contribution of both $s$ and $p$ polarizations. In the above-gap range, both G-FBN-G-InSb cell and G-FBN-G-FBN-InSb cell exhibit enhanced photon transmission, thanks to the SPPPs supported by the graphene-{\it h}-BN heterostructures. Fig.~4 shows that the photon transmission spectra of the two near-field NTPV systems do not differ much. Therefore, the performances of the two NTPV systems are comparable.

\begin{figure}
\centering\includegraphics[width=12cm]{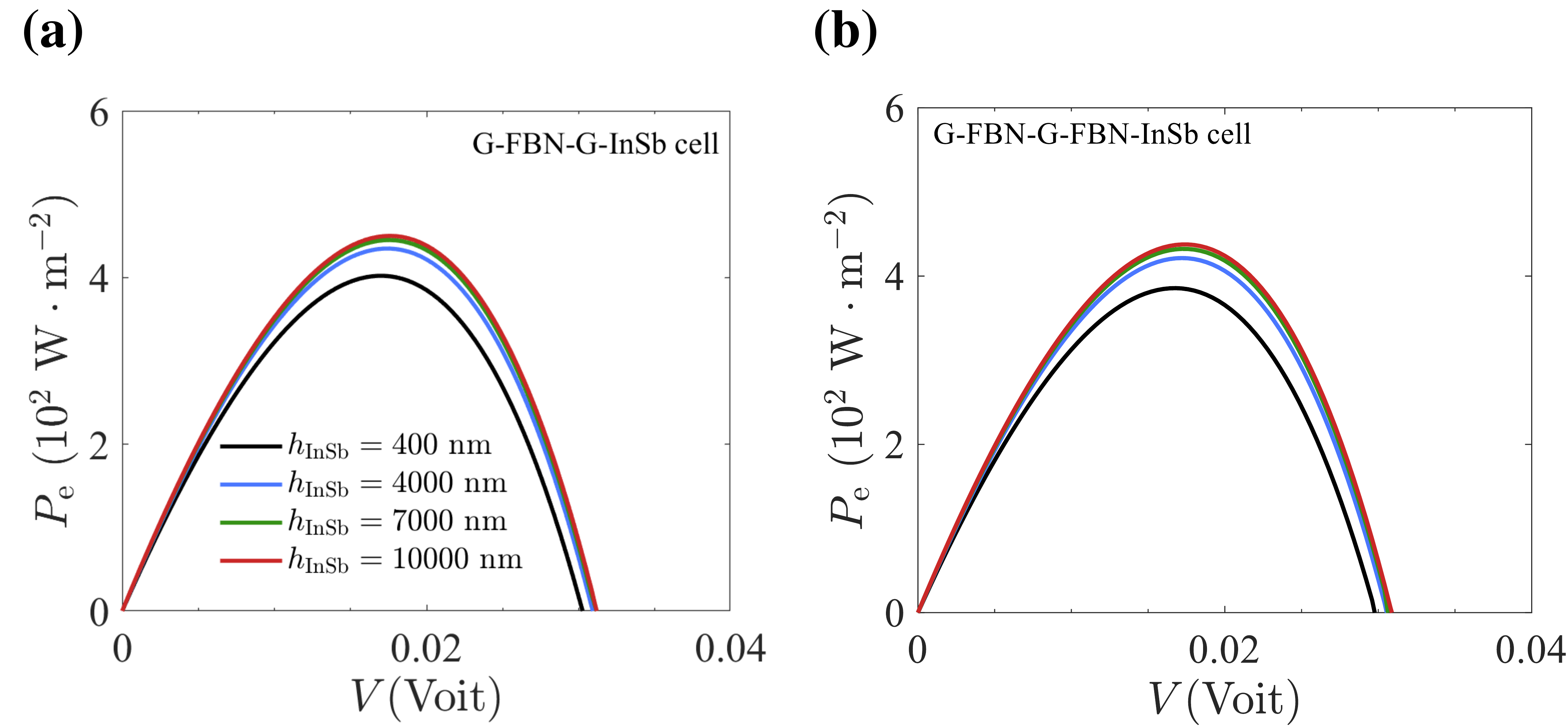}
\centering\includegraphics[width=12cm]{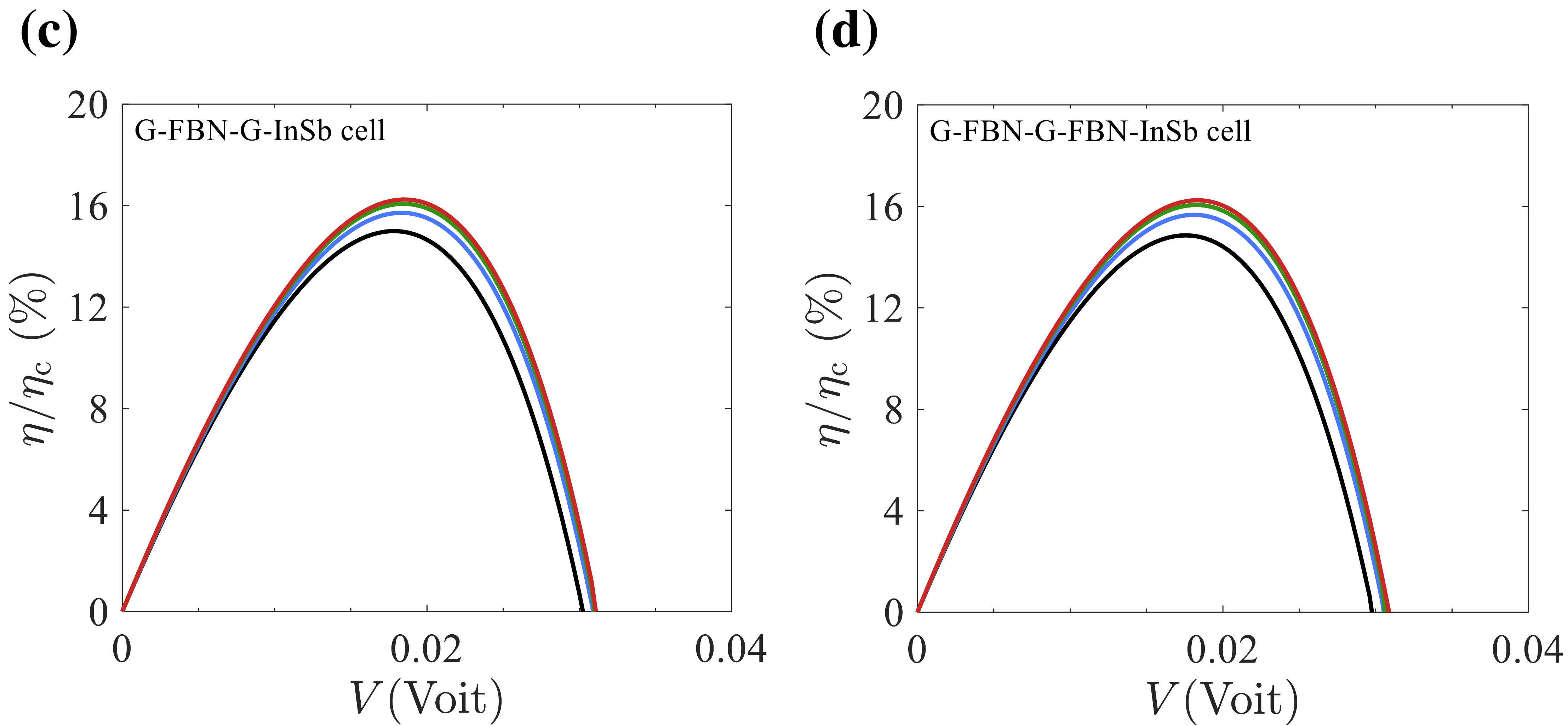}
\caption{Performances of two near-field TPV cells for various InSb thin-film thickness. (a) Electrical power density and (b) energy$-$conversion efficiency in units of Carnot efficiency ($\eta_{\rm c}$). The temperatures of the emitter and the cell are set at $T_{\rm emit}=450$ K and $T_{\rm cell}=320$ K, respectively. The vacuum gap distance is $d=20$ nm, the {\it h}-BN film thickness is $h_{\rm BN}=20$ nm and the chemical potential of graphene is $\mu_{\rm g}=1.0$ eV.}\label{fig:performance_hInSb}
\end{figure}

\begin{figure}
\centering\includegraphics[width=12cm]{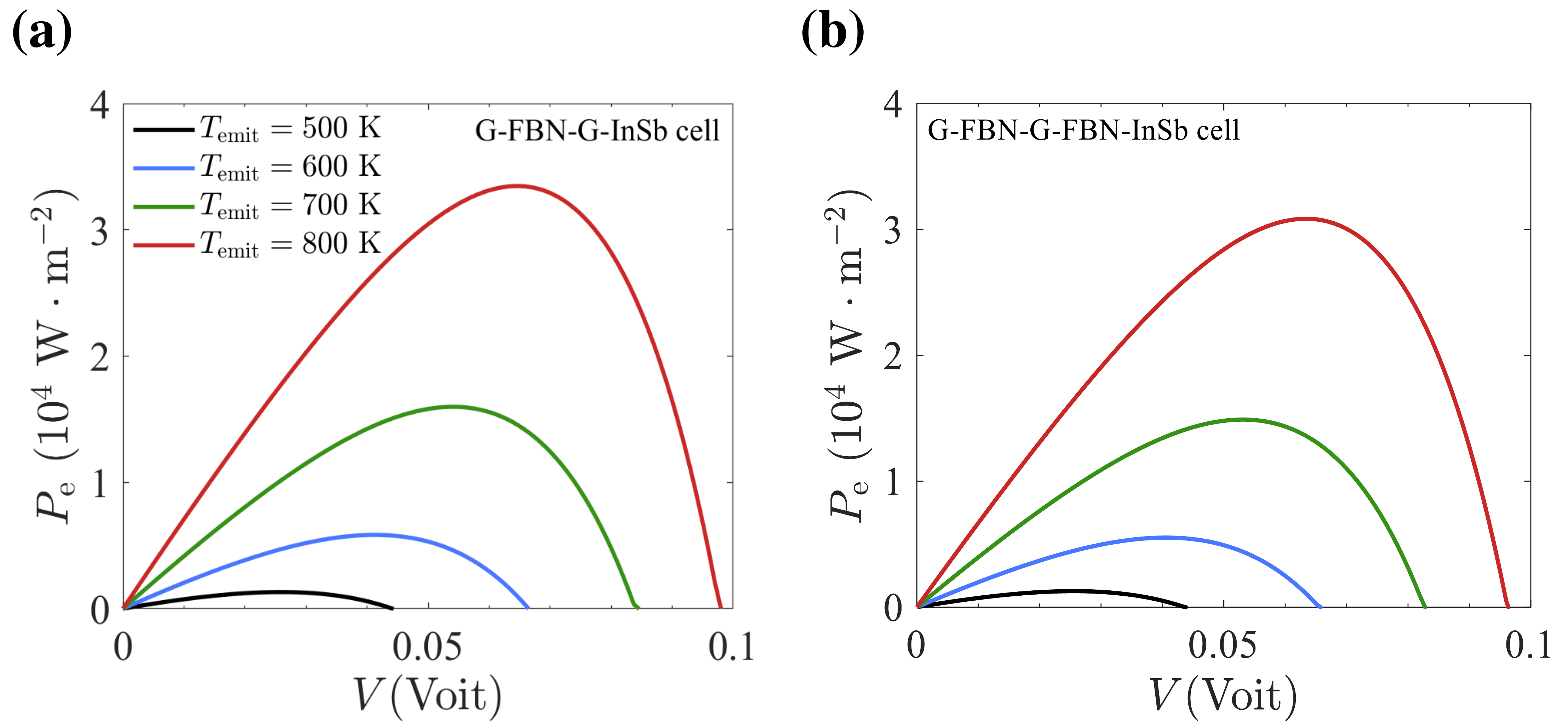}
\centering\includegraphics[width=12cm]{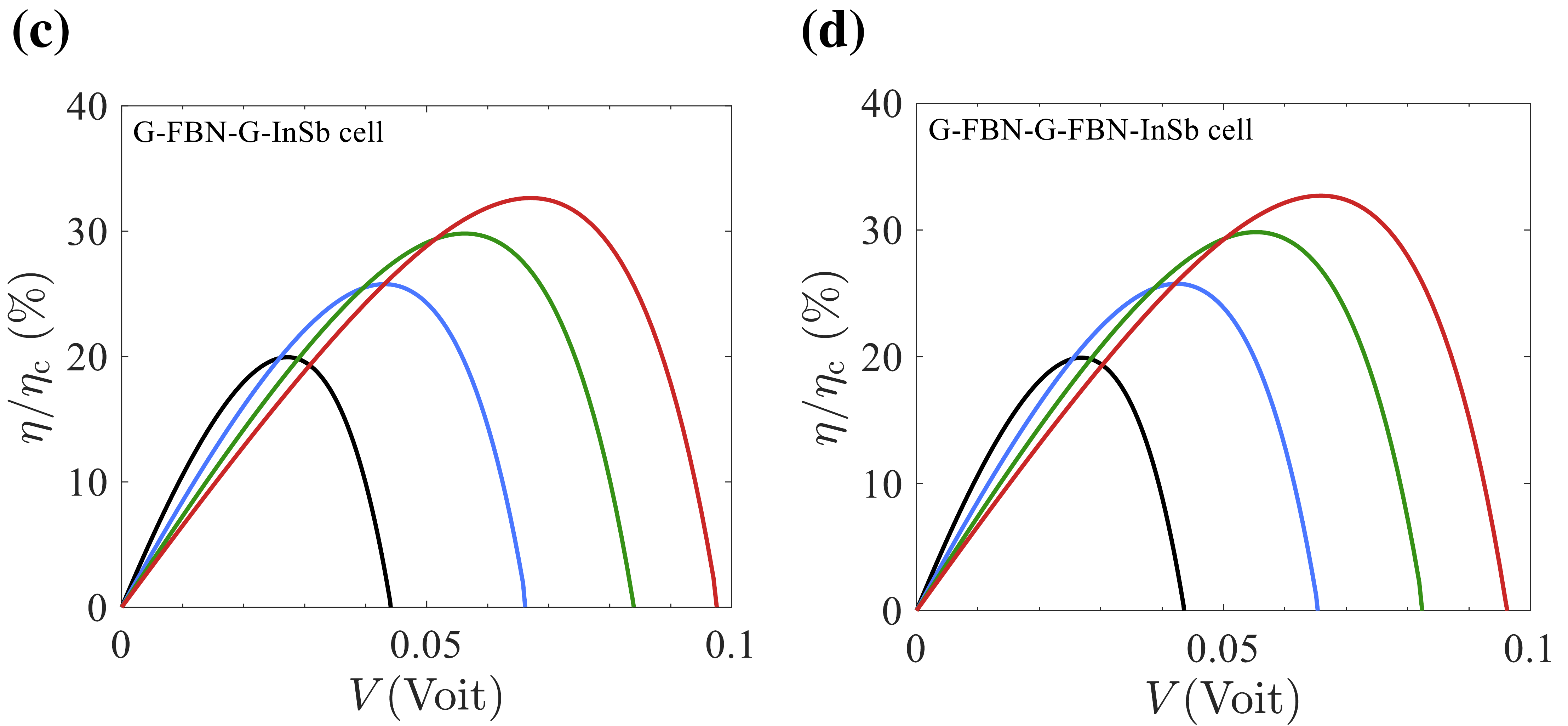}
\caption{Performances of two near-field TPV cells for various emitter temperatures. (a) Electrical power density and (b) energy$-$conversion efficiency in units of Carnot efficiency ($\eta_{\rm c}$). The temperature of  the cell is set at $T_{\rm cell}=320$ K. The vacuum gap distance is $d=20$ nm, the {\it h}-BN film thickness is $h_{\rm BN}=20$ nm and the thickness of the InSb thin film is chosen as an optimal value of $h_{\rm InSb}=10000$ nm. The chemical potential of graphene is $\mu_{\rm g}=1.0$ eV.}\label{fig:performance_Temit}
\end{figure}

\section{Optimization of graphene-{\it h}-BN-InSb near-field systems}\label{nfoptimization}

\begin{figure}
	\centering\includegraphics[width=12cm]{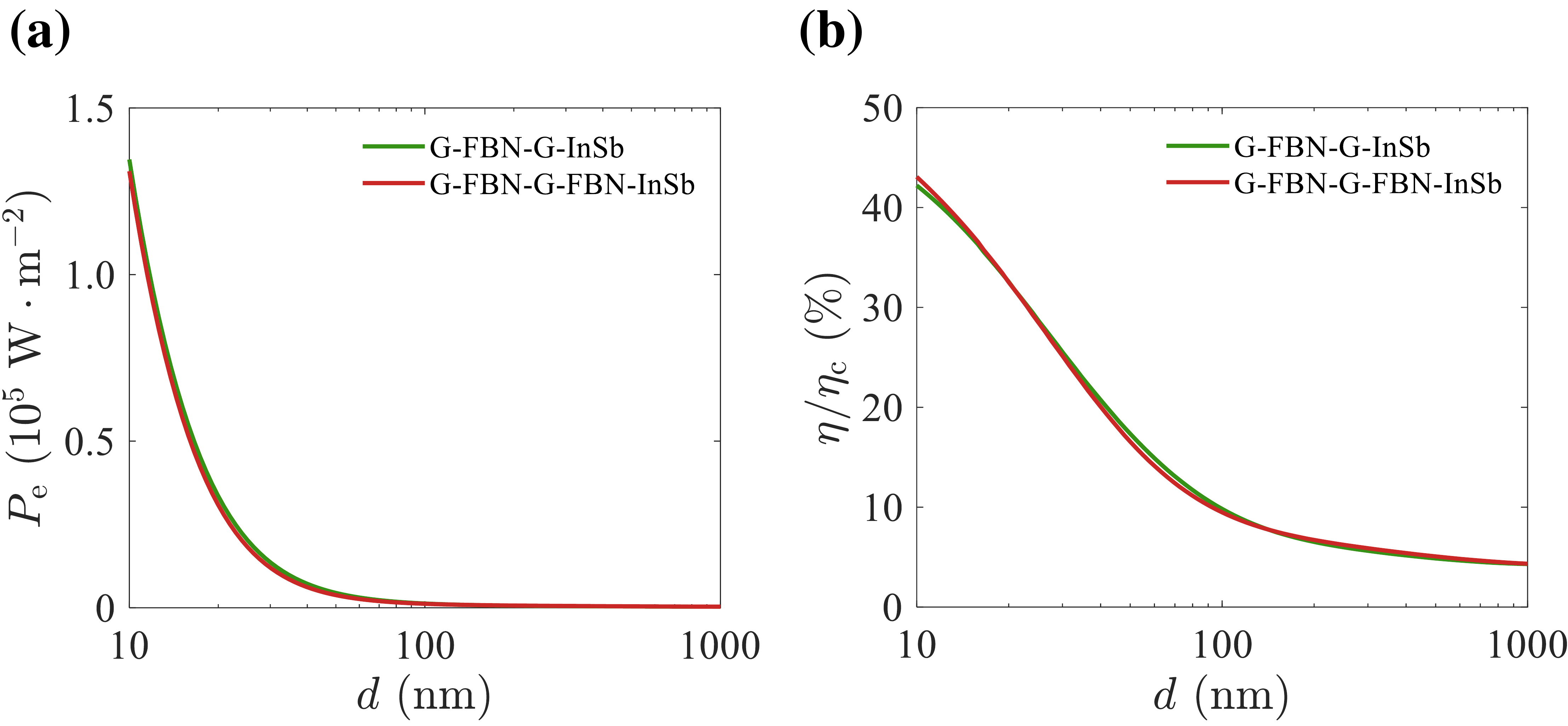}
	\caption{Optimal performances of two NTPV cells. (a) Maximum  electrical power density and (b) energy$-$conversion efficiency at maximum electric power in units of Carnot efficiency ($\eta_{\rm c}$). The temperatures of the emitter and the cell are set at $T_{\rm emit}=800$ K and $T_{\rm cell}=320$ K, respectively.
		The {\it h}-BN film thickness is $h_{\rm BN}=20$ nm and the thickness of the InSb thin film is $h_{\rm InSb}=10000$ nm. The chemical potential of graphene is $\mu_{\rm g}=1.0$ eV. The Carnot efficiency is given by $\eta_{\rm c}=1-{T_{\rm cell}}/{T_{\rm emit}}$.}\label{fig:optimal performance}
\end{figure}

We now study the performances of the two near-field systems for various InSb thin-film thicknesses and emitter temperatures in Figs.~\ref{fig:performance_hInSb} and~\ref{fig:performance_Temit}. As presented in Fig.~\ref{fig:performance_hInSb} that both of the output power density and the energy-conversion efficiency of the two setups are improved with increasing InSb thin-film thickness. Especially for $h_{\rm InSb}=10000$ nm, the maximum power densities of the G-FBN-G-InSb cell and G-FBN-G-FBN-InSb cell are about $4.5\times10^{2}~\rm{W/m^2}$ and $4.0\times10^{2}~\rm{W/m^2}$, respectively. The maximum values of the corresponding efficiency are close to $16\%\eta_{\rm C}$ and  $15\%\eta_{\rm C}$, respectively. This enhancement can be attributed to the increased absorption factor $\left(1-\exp{\left[-\rm Imag(\it k_{z}^{\rm InSb})\it h_{\rm InSb}\right]}\right)$ because it is dependent on $h_{\rm InSb}$. However, when the InSb thin-film thickness is increased from 4000 nm to 10000 nm, the increase of the output power density and the energy-conversion efficiency for both systems are soon saturated. This is essentially due to the near-field nature of the heat transfer: the photons transferred from the emitter to the InSb cell is dominated by the evanescent electromagnetic waves which are highly localized at the surface of the emitter. Further increase of the thickness of the InSb cell does not change the amount of photons received by the InSb cell.

Fig.~\ref{fig:performance_Temit} shows that by increasing the emitter temperature, both the performances of the two setups can be improved by orders of magnitudes. Here, the InSb thin-film thickness has been chosen as the optimal value with $h_{\rm InSb}=10000$ nm. For the G-FBN-G-InSb cell, the maximum output power density and energy efficiency at $T_{\rm emit}=800$ K are increased to $3.3\times10^{4}~\rm{W/m^2}$ and $33\%\eta_c$, respectively.
For the G-FBN-G-FBN-InSb cell, the maximum output power density and energy efficiency at $T_{\rm emit}=800$ K are also improved to $3.1\times10^{4}~\rm{W/m^2}$ and $32\%\eta_c$, respectively. 

We now examine the maximum output electric power density and the efficiency at maximum power. We consider specifically the situation with $h_{\rm InSb}=10000$~nm and $T_{\rm emit}=800$~K. As shown in Fig.~\ref{fig:optimal performance}, both the maximum electric power density and the efficiency at the maximum power of the two near-field NTPV systems dramatically increase as the vacuum gap $d$ decreases. The best performances are achieved when the vacuum gap is $d=10$ nm for both systems. Specifically, the maximum electric power density reaches to $1.3\times10^{5}~\rm{W/m^2}$, while the efficiency at maximum power goes up to $42\%$ of the Carnot efficiency. The two near-field NTPV systems perform nearly equally well.

\section{Summary and conclusions}\label{conclusions}

In conclusion, we investigated two NTPV devices, denoted as the G-FBN-G-InSb cell and G-FBN-G-FBN-InSb cell, which have different emitters. The purpose of the investigation is to find a high-performance emitter design with relatively simple structure. Indeed, we find that both systems perform better than the near-field NTPV system based on mono graphene-$h$-BN heterostructure, i.e., graphene-$h$-BN-InSb cell~\cite{PRAppliedWang}. The optimal output power density of the G-FBN-G-InSb cell can reach to $1.3\times10^{5}~\rm{W\cdot m^{-2}}$, nearly twice of the optimal power density of the graphene-$h$-BN-InSb cell. The optimal energy efficiency can be as large as $42\%$ of the Carnot efficiency, which is $24\%$ higher than the optimal efficency of the monocell-system. 
We analyze the underlying physical mechanisms that lead to the excellent performances of the G-FBN-G-InSb cell. We further show that the performance of the proposed NTPV system is affected negligibly by the finite thickness of the InSb cell which is due to the near-field nature of the heat transfer: the absorbed photons are highly localized at the surface of the InSb cell. Our study shows that NTPV systems are promising for high-performance moderate temperature thermal-to-electric energy conversion.

\section{Acknowledgements}
We acknowledge support from the National Natural Science Foundation of China (NSFC Grant No. 11675116), the National Natural Science Foundation of China (NSFC Grant No. 12074281), the Jiangsu distinguished professor funding and a Project Funded by the Priority Academic Program Development of Jiangsu Higher Education Institutions (PAPD). J.L thanks China's Postdoctoral Science Foundation funded project (No. 2020M681376).

\bibliography{ref_thInSb}

\begin{thebibliography}{44}%
\makeatletter
\providecommand \@ifxundefined [1]{%
 \@ifx{#1\undefined}
}%
\providecommand \@ifnum [1]{%
 \ifnum #1\expandafter \@firstoftwo
 \else \expandafter \@secondoftwo
 \fi
}%
\providecommand \@ifx [1]{%
 \ifx #1\expandafter \@firstoftwo
 \else \expandafter \@secondoftwo
 \fi
}%
\providecommand \natexlab [1]{#1}%
\providecommand \enquote  [1]{``#1''}%
\providecommand \bibnamefont  [1]{#1}%
\providecommand \bibfnamefont [1]{#1}%
\providecommand \citenamefont [1]{#1}%
\providecommand \href@noop [0]{\@secondoftwo}%
\providecommand \href [0]{\begingroup \@sanitize@url \@href}%
\providecommand \@href[1]{\@@startlink{#1}\@@href}%
\providecommand \@@href[1]{\endgroup#1\@@endlink}%
\providecommand \@sanitize@url [0]{\catcode `\\12\catcode `\$12\catcode
  `\&12\catcode `\#12\catcode `\^12\catcode `\_12\catcode `\%12\relax}%
\providecommand \@@startlink[1]{}%
\providecommand \@@endlink[0]{}%
\providecommand \url  [0]{\begingroup\@sanitize@url \@url }%
\providecommand \@url [1]{\endgroup\@href {#1}{\urlprefix }}%
\providecommand \urlprefix  [0]{URL }%
\providecommand \Eprint [0]{\href }%
\providecommand \doibase [0]{http://dx.doi.org/}%
\providecommand \selectlanguage [0]{\@gobble}%
\providecommand \bibinfo  [0]{\@secondoftwo}%
\providecommand \bibfield  [0]{\@secondoftwo}%
\providecommand \translation [1]{[#1]}%
\providecommand \BibitemOpen [0]{}%
\providecommand \bibitemStop [0]{}%
\providecommand \bibitemNoStop [0]{.\EOS\space}%
\providecommand \EOS [0]{\spacefactor3000\relax}%
\providecommand \BibitemShut  [1]{\csname bibitem#1\endcsname}%
\let\auto@bib@innerbib\@empty
\bibitem [{\citenamefont {Shockley}\ and\ \citenamefont
  {Queisser}(1961)}]{shockley1961detailed}%
  \BibitemOpen
  \bibfield  {author} {\bibinfo {author} {\bibfnamefont {W.}~\bibnamefont
  {Shockley}}\ and\ \bibinfo {author} {\bibfnamefont {H.~J.}\ \bibnamefont
  {Queisser}},\ }\bibfield  {title} {\enquote {\bibinfo {title} {Detailed
  balance limit of efficiency of p-n junction solar cells},}\ }\href@noop {}
  {\bibfield  {journal} {\bibinfo  {journal} {J. Appl. Phys.}\ }\textbf
  {\bibinfo {volume} {32}},\ \bibinfo {pages} {510--519} (\bibinfo {year}
  {1961})}\BibitemShut {NoStop}%
\bibitem [{\citenamefont {Mart{\'\i}n}\ and\ \citenamefont
  {Algora}(2004)}]{martin2004temperature}%
  \BibitemOpen
  \bibfield  {author} {\bibinfo {author} {\bibfnamefont {D.}~\bibnamefont
  {Mart{\'\i}n}}\ and\ \bibinfo {author} {\bibfnamefont {C.}~\bibnamefont
  {Algora}},\ }\bibfield  {title} {\enquote {\bibinfo {title}
  {Temperature-dependent gasb material parameters for reliable
  thermophotovoltaic cell modelling},}\ }\href@noop {} {\bibfield  {journal}
  {\bibinfo  {journal} {Semicond. Sci. Tech.}\ }\textbf {\bibinfo {volume}
  {19}},\ \bibinfo {pages} {1040} (\bibinfo {year} {2004})}\BibitemShut
  {NoStop}%
\bibitem [{\citenamefont {Nagashima}\ \emph {et~al.}(2007)\citenamefont
  {Nagashima}, \citenamefont {Okumura},\ and\ \citenamefont
  {Yamaguchi}}]{nagashima2007germanium}%
  \BibitemOpen
  \bibfield  {author} {\bibinfo {author} {\bibfnamefont {T.}~\bibnamefont
  {Nagashima}}, \bibinfo {author} {\bibfnamefont {K.}~\bibnamefont {Okumura}},
  \ and\ \bibinfo {author} {\bibfnamefont {M.}~\bibnamefont {Yamaguchi}},\
  }\bibfield  {title} {\enquote {\bibinfo {title} {A germanium back contact
  type thermophotovoltaic cell},}\ }\href@noop {} {\bibfield  {journal}
  {\bibinfo  {journal} {AIP Conf. Proc.}\ }\textbf {\bibinfo {volume} {890}},\
  \bibinfo {pages} {174--181} (\bibinfo {year} {2007})}\BibitemShut {NoStop}%
\bibitem [{\citenamefont {Fraas}\ and\ \citenamefont
  {Ferguson}(2000)}]{fraas2000three}%
  \BibitemOpen
  \bibfield  {author} {\bibinfo {author} {\bibfnamefont {L.~M.}\ \bibnamefont
  {Fraas}}\ and\ \bibinfo {author} {\bibfnamefont {L.~G.}\ \bibnamefont
  {Ferguson}},\ }\bibfield  {title} {\enquote {\bibinfo {title} {Three-layer
  solid infrared emitter with spectral output matched to low bandgap
  thermophotovoltaic cells},}\ }\href@noop {} {\bibfield  {journal} {\bibinfo
  {journal} {US Patent 6, 091, 018}\ } (\bibinfo {year} {2000})}\BibitemShut
  {NoStop}%
\bibitem [{\citenamefont {Sulima}\ and\ \citenamefont
  {Bett}(2001)}]{sulima2001fabrication}%
  \BibitemOpen
  \bibfield  {author} {\bibinfo {author} {\bibfnamefont {O.~V.}\ \bibnamefont
  {Sulima}}\ and\ \bibinfo {author} {\bibfnamefont {A.~W.}\ \bibnamefont
  {Bett}},\ }\bibfield  {title} {\enquote {\bibinfo {title} {Fabrication and
  simulation of $\rm{GaSb}$ thermophotovoltaic cells},}\ }\href@noop {}
  {\bibfield  {journal} {\bibinfo  {journal} {Sol. Energ. Mat. Sol. C.}\
  }\textbf {\bibinfo {volume} {66}},\ \bibinfo {pages} {533--540} (\bibinfo
  {year} {2001})}\BibitemShut {NoStop}%
\bibitem [{\citenamefont {Wu}\ \emph {et~al.}(2012)\citenamefont {Wu},
  \citenamefont {Neuner~III}, \citenamefont {John}, \citenamefont {Milder},
  \citenamefont {Zollars}, \citenamefont {Savoy},\ and\ \citenamefont
  {Shvets}}]{wu2012metamaterial}%
  \BibitemOpen
  \bibfield  {author} {\bibinfo {author} {\bibfnamefont {C.}~\bibnamefont
  {Wu}}, \bibinfo {author} {\bibfnamefont {B.}~\bibnamefont {Neuner~III}},
  \bibinfo {author} {\bibfnamefont {J.}~\bibnamefont {John}}, \bibinfo {author}
  {\bibfnamefont {A.}~\bibnamefont {Milder}}, \bibinfo {author} {\bibfnamefont
  {B.}~\bibnamefont {Zollars}}, \bibinfo {author} {\bibfnamefont
  {S.}~\bibnamefont {Savoy}}, \ and\ \bibinfo {author} {\bibfnamefont
  {G.}~\bibnamefont {Shvets}},\ }\bibfield  {title} {\enquote {\bibinfo {title}
  {Metamaterial-based integrated plasmonic absorber/emitter for solar
  thermo-photovoltaic systems},}\ }\href@noop {} {\bibfield  {journal}
  {\bibinfo  {journal} {J. Optics}\ }\textbf {\bibinfo {volume} {14}},\
  \bibinfo {pages} {024005} (\bibinfo {year} {2012})}\BibitemShut {NoStop}%
\bibitem [{\citenamefont {Chan}\ \emph {et~al.}(2013)\citenamefont {Chan},
  \citenamefont {Bermel}, \citenamefont {Pilawa-Podgurski}, \citenamefont
  {Marton}, \citenamefont {Jensen}, \citenamefont {Senkevich}, \citenamefont
  {Joannopoulos}, \citenamefont {Solja{\v{c}}i{\'c}},\ and\ \citenamefont
  {Celanovic}}]{chan2013toward}%
  \BibitemOpen
  \bibfield  {author} {\bibinfo {author} {\bibfnamefont {W.~R.}\ \bibnamefont
  {Chan}}, \bibinfo {author} {\bibfnamefont {P.}~\bibnamefont {Bermel}},
  \bibinfo {author} {\bibfnamefont {R.~C.~N.}\ \bibnamefont
  {Pilawa-Podgurski}}, \bibinfo {author} {\bibfnamefont {C.~H}\ \bibnamefont
  {Marton}}, \bibinfo {author} {\bibfnamefont {K.~F.}\ \bibnamefont {Jensen}},
  \bibinfo {author} {\bibfnamefont {J.~J.}\ \bibnamefont {Senkevich}}, \bibinfo
  {author} {\bibfnamefont {J.~D.}\ \bibnamefont {Joannopoulos}}, \bibinfo
  {author} {\bibfnamefont {M.}~\bibnamefont {Solja{\v{c}}i{\'c}}}, \ and\
  \bibinfo {author} {\bibfnamefont {I.}~\bibnamefont {Celanovic}},\ }\bibfield
  {title} {\enquote {\bibinfo {title} {Toward high-energy-density,
  high-efficiency, and moderate-temperature chip-scale thermophotovoltaics},}\
  }\href@noop {} {\bibfield  {journal} {\bibinfo  {journal} {Proc. Nat. Acad.
  Sci.}\ }\textbf {\bibinfo {volume} {110}},\ \bibinfo {pages} {5309--5314}
  (\bibinfo {year} {2013})}\BibitemShut {NoStop}%
\bibitem [{\citenamefont {Liao}\ \emph {et~al.}(2016)\citenamefont {Liao},
  \citenamefont {Cai}, \citenamefont {Zhao},\ and\ \citenamefont
  {Chen}}]{liao2016efficiently}%
  \BibitemOpen
  \bibfield  {author} {\bibinfo {author} {\bibfnamefont {T.}~\bibnamefont
  {Liao}}, \bibinfo {author} {\bibfnamefont {L.}~\bibnamefont {Cai}}, \bibinfo
  {author} {\bibfnamefont {Y.}~\bibnamefont {Zhao}}, \ and\ \bibinfo {author}
  {\bibfnamefont {J.}~\bibnamefont {Chen}},\ }\bibfield  {title} {\enquote
  {\bibinfo {title} {Efficiently exploiting the waste heat in solid oxide fuel
  cell by means of thermophotovoltaic cell},}\ }\href@noop {} {\bibfield
  {journal} {\bibinfo  {journal} {J. Power Sources}\ }\textbf {\bibinfo
  {volume} {306}},\ \bibinfo {pages} {666--673} (\bibinfo {year}
  {2016})}\BibitemShut {NoStop}%
\bibitem [{\citenamefont {Zhao}\ \emph
  {et~al.}(2017{\natexlab{a}})\citenamefont {Zhao}, \citenamefont {Chen},
  \citenamefont {Buddhiraju}, \citenamefont {Bhatt}, \citenamefont {Lipson},\
  and\ \citenamefont {Fan}}]{zhao2017high}%
  \BibitemOpen
  \bibfield  {author} {\bibinfo {author} {\bibfnamefont {B.}~\bibnamefont
  {Zhao}}, \bibinfo {author} {\bibfnamefont {K.}~\bibnamefont {Chen}}, \bibinfo
  {author} {\bibfnamefont {S.}~\bibnamefont {Buddhiraju}}, \bibinfo {author}
  {\bibfnamefont {G.}~\bibnamefont {Bhatt}}, \bibinfo {author} {\bibfnamefont
  {M.}~\bibnamefont {Lipson}}, \ and\ \bibinfo {author} {\bibfnamefont
  {S.}~\bibnamefont {Fan}},\ }\bibfield  {title} {\enquote {\bibinfo {title}
  {High-performance near-field thermophotovoltaics for waste heat recovery},}\
  }\href@noop {} {\bibfield  {journal} {\bibinfo  {journal} {Nano Energy}\
  }\textbf {\bibinfo {volume} {41}},\ \bibinfo {pages} {344--350} (\bibinfo
  {year} {2017}{\natexlab{a}})}\BibitemShut {NoStop}%
\bibitem [{\citenamefont {Tervo}\ \emph {et~al.}(2018)\citenamefont {Tervo},
  \citenamefont {Bagherisereshki},\ and\ \citenamefont {Zhang}}]{Tervo2018}%
  \BibitemOpen
  \bibfield  {author} {\bibinfo {author} {\bibfnamefont {E.}~\bibnamefont
  {Tervo}}, \bibinfo {author} {\bibfnamefont {E.}~\bibnamefont
  {Bagherisereshki}}, \ and\ \bibinfo {author} {\bibfnamefont {Z.~M.}\
  \bibnamefont {Zhang}},\ }\bibfield  {title} {\enquote {\bibinfo {title}
  {Near-field radiative thermoelectric energy converters: a review},}\
  }\href@noop {} {\bibfield  {journal} {\bibinfo  {journal} {Front. Energy}\
  }\textbf {\bibinfo {volume} {12}},\ \bibinfo {pages} {5--21} (\bibinfo {year}
  {2018})}\BibitemShut {NoStop}%
\bibitem [{\citenamefont {Svetovoy}\ \emph {et~al.}(2012)\citenamefont
  {Svetovoy}, \citenamefont {Van~Zwol},\ and\ \citenamefont
  {Chevrier}}]{svetovoy2012plasmon}%
  \BibitemOpen
  \bibfield  {author} {\bibinfo {author} {\bibfnamefont {V.~B.}\ \bibnamefont
  {Svetovoy}}, \bibinfo {author} {\bibfnamefont {P.~J.}\ \bibnamefont
  {Van~Zwol}}, \ and\ \bibinfo {author} {\bibfnamefont {J.}~\bibnamefont
  {Chevrier}},\ }\bibfield  {title} {\enquote {\bibinfo {title} {Plasmon
  enhanced near-field radiative heat transfer for graphene covered
  dielectrics},}\ }\href@noop {} {\bibfield  {journal} {\bibinfo  {journal}
  {Phys. Rev. B}\ }\textbf {\bibinfo {volume} {85}},\ \bibinfo {pages} {155418}
  (\bibinfo {year} {2012})}\BibitemShut {NoStop}%
\bibitem [{\citenamefont {Ilic}\ \emph {et~al.}(2012)\citenamefont {Ilic},
  \citenamefont {Jablan}, \citenamefont {Joannopoulos}, \citenamefont
  {Celanovic},\ and\ \citenamefont {Solja{\v{c}}i{\'c}}}]{ilic2012overcoming}%
  \BibitemOpen
  \bibfield  {author} {\bibinfo {author} {\bibfnamefont {O.}~\bibnamefont
  {Ilic}}, \bibinfo {author} {\bibfnamefont {M.}~\bibnamefont {Jablan}},
  \bibinfo {author} {\bibfnamefont {J.~D.}\ \bibnamefont {Joannopoulos}},
  \bibinfo {author} {\bibfnamefont {I.}~\bibnamefont {Celanovic}}, \ and\
  \bibinfo {author} {\bibfnamefont {M.}~\bibnamefont {Solja{\v{c}}i{\'c}}},\
  }\bibfield  {title} {\enquote {\bibinfo {title} {Overcoming the black body
  limit in plasmonic and graphene near-field thermophotovoltaic systems},}\
  }\href@noop {} {\bibfield  {journal} {\bibinfo  {journal} {Opt. Express}\
  }\textbf {\bibinfo {volume} {20}},\ \bibinfo {pages} {A366--A384} (\bibinfo
  {year} {2012})}\BibitemShut {NoStop}%
\bibitem [{\citenamefont {Svetovoy}\ and\ \citenamefont
  {Palasantzas}(2014)}]{svetovoy2014graphene}%
  \BibitemOpen
  \bibfield  {author} {\bibinfo {author} {\bibfnamefont {V.~B.}\ \bibnamefont
  {Svetovoy}}\ and\ \bibinfo {author} {\bibfnamefont {G.}~\bibnamefont
  {Palasantzas}},\ }\bibfield  {title} {\enquote {\bibinfo {title}
  {Graphene-on-silicon near-field thermophotovoltaic cell},}\ }\href@noop {}
  {\bibfield  {journal} {\bibinfo  {journal} {Phys. Rev. Appl.}\ }\textbf
  {\bibinfo {volume} {2}},\ \bibinfo {pages} {034006} (\bibinfo {year}
  {2014})}\BibitemShut {NoStop}%
\bibitem [{\citenamefont {Basu}\ \emph {et~al.}(2015)\citenamefont {Basu},
  \citenamefont {Yang},\ and\ \citenamefont {Wang}}]{basu2015near}%
  \BibitemOpen
  \bibfield  {author} {\bibinfo {author} {\bibfnamefont {S.}~\bibnamefont
  {Basu}}, \bibinfo {author} {\bibfnamefont {Y.}~\bibnamefont {Yang}}, \ and\
  \bibinfo {author} {\bibfnamefont {L.}~\bibnamefont {Wang}},\ }\bibfield
  {title} {\enquote {\bibinfo {title} {Near-field radiative heat transfer
  between metamaterials coated with silicon carbide thin films},}\ }\href@noop
  {} {\bibfield  {journal} {\bibinfo  {journal} {Appl. Phys. Lett.}\ }\textbf
  {\bibinfo {volume} {106}},\ \bibinfo {pages} {033106} (\bibinfo {year}
  {2015})}\BibitemShut {NoStop}%
\bibitem [{\citenamefont {Narayanaswamy}\ and\ \citenamefont
  {Chen}(2003)}]{narayanaswamy2003surface}%
  \BibitemOpen
  \bibfield  {author} {\bibinfo {author} {\bibfnamefont {A.}~\bibnamefont
  {Narayanaswamy}}\ and\ \bibinfo {author} {\bibfnamefont {G.}~\bibnamefont
  {Chen}},\ }\bibfield  {title} {\enquote {\bibinfo {title} {Surface modes for
  near field thermophotovoltaics},}\ }\href@noop {} {\bibfield  {journal}
  {\bibinfo  {journal} {Appl. Phys. Lett.}\ }\textbf {\bibinfo {volume} {82}},\
  \bibinfo {pages} {3544--3546} (\bibinfo {year} {2003})}\BibitemShut {NoStop}%
\bibitem [{\citenamefont {Laroche}\ \emph {et~al.}(2006)\citenamefont
  {Laroche}, \citenamefont {Carminati},\ and\ \citenamefont
  {Greffet}}]{laroche2006JAP}%
  \BibitemOpen
  \bibfield  {author} {\bibinfo {author} {\bibfnamefont {M.}~\bibnamefont
  {Laroche}}, \bibinfo {author} {\bibfnamefont {R.}~\bibnamefont {Carminati}},
  \ and\ \bibinfo {author} {\bibfnamefont {J.~J.}\ \bibnamefont {Greffet}},\
  }\bibfield  {title} {\enquote {\bibinfo {title} {Near-field
  thermophotovoltaic energy conversion},}\ }\href@noop {} {\bibfield  {journal}
  {\bibinfo  {journal} {J. Appl. Phys.}\ }\textbf {\bibinfo {volume} {100}},\
  \bibinfo {pages} {063704} (\bibinfo {year} {2006})}\BibitemShut {NoStop}%
\bibitem [{\citenamefont {Park}\ \emph {et~al.}(2008)\citenamefont {Park},
  \citenamefont {Basu}, \citenamefont {King},\ and\ \citenamefont
  {Zhang}}]{park2008performance}%
  \BibitemOpen
  \bibfield  {author} {\bibinfo {author} {\bibfnamefont {K.}~\bibnamefont
  {Park}}, \bibinfo {author} {\bibfnamefont {S.}~\bibnamefont {Basu}}, \bibinfo
  {author} {\bibfnamefont {W.~P.}\ \bibnamefont {King}}, \ and\ \bibinfo
  {author} {\bibfnamefont {Z.~M.}\ \bibnamefont {Zhang}},\ }\bibfield  {title}
  {\enquote {\bibinfo {title} {Performance analysis of near-field
  thermophotovoltaic devices considering absorption distribution},}\
  }\href@noop {} {\bibfield  {journal} {\bibinfo  {journal} {J. Quant.
  Spectros. Radiat. Transfer}\ }\textbf {\bibinfo {volume} {109}},\ \bibinfo
  {pages} {305--316} (\bibinfo {year} {2008})}\BibitemShut {NoStop}%
\bibitem [{\citenamefont {Bright}\ \emph {et~al.}(2014)\citenamefont {Bright},
  \citenamefont {Wang},\ and\ \citenamefont {Zhang}}]{bright2014performance}%
  \BibitemOpen
  \bibfield  {author} {\bibinfo {author} {\bibfnamefont {T.~J.}\ \bibnamefont
  {Bright}}, \bibinfo {author} {\bibfnamefont {L.~P.}\ \bibnamefont {Wang}}, \
  and\ \bibinfo {author} {\bibfnamefont {Z.~M.}\ \bibnamefont {Zhang}},\
  }\bibfield  {title} {\enquote {\bibinfo {title} {Performance of near-field
  thermophotovoltaic cells enhanced with a backside reflector},}\ }\href@noop
  {} {\bibfield  {journal} {\bibinfo  {journal} {J. Heat Transfer}\ }\textbf
  {\bibinfo {volume} {136}},\ \bibinfo {pages} {062701} (\bibinfo {year}
  {2014})}\BibitemShut {NoStop}%
\bibitem [{\citenamefont {Molesky}\ and\ \citenamefont
  {Jacob}(2015)}]{molesky2015ideal}%
  \BibitemOpen
  \bibfield  {author} {\bibinfo {author} {\bibfnamefont {S.}~\bibnamefont
  {Molesky}}\ and\ \bibinfo {author} {\bibfnamefont {Z.}~\bibnamefont
  {Jacob}},\ }\bibfield  {title} {\enquote {\bibinfo {title} {Ideal near-field
  thermophotovoltaic cells},}\ }\href@noop {} {\bibfield  {journal} {\bibinfo
  {journal} {Phys. Rev. B}\ }\textbf {\bibinfo {volume} {91}},\ \bibinfo
  {pages} {205435} (\bibinfo {year} {2015})}\BibitemShut {NoStop}%
\bibitem [{\citenamefont {St-Gelais}\ \emph {et~al.}(2017)\citenamefont
  {St-Gelais}, \citenamefont {Bhatt}, \citenamefont {Zhu}, \citenamefont
  {Fan},\ and\ \citenamefont {Lipson}}]{st2017hot}%
  \BibitemOpen
  \bibfield  {author} {\bibinfo {author} {\bibfnamefont {R.}~\bibnamefont
  {St-Gelais}}, \bibinfo {author} {\bibfnamefont {G.~R.}\ \bibnamefont
  {Bhatt}}, \bibinfo {author} {\bibfnamefont {L.}~\bibnamefont {Zhu}}, \bibinfo
  {author} {\bibfnamefont {S.}~\bibnamefont {Fan}}, \ and\ \bibinfo {author}
  {\bibfnamefont {M.}~\bibnamefont {Lipson}},\ }\bibfield  {title} {\enquote
  {\bibinfo {title} {Hot carrier-based near-field thermophotovoltaic energy
  conversion},}\ }\href@noop {} {\bibfield  {journal} {\bibinfo  {journal} {ACS
  nano}\ }\textbf {\bibinfo {volume} {11}},\ \bibinfo {pages} {3001--3009}
  (\bibinfo {year} {2017})}\BibitemShut {NoStop}%
\bibitem [{\citenamefont {Jiang}\ and\ \citenamefont
  {Imry}(2018)}]{Jiang2018Near}%
  \BibitemOpen
  \bibfield  {author} {\bibinfo {author} {\bibfnamefont {J.-H.}\ \bibnamefont
  {Jiang}}\ and\ \bibinfo {author} {\bibfnamefont {Yoseph}\ \bibnamefont
  {Imry}},\ }\bibfield  {title} {\enquote {\bibinfo {title} {Near-field
  three-terminal thermoelectric heat engine},}\ }\href@noop {} {\bibfield
  {journal} {\bibinfo  {journal} {Phys. Rev. B}\ }\textbf {\bibinfo {volume}
  {97}} (\bibinfo {year} {2018})}\BibitemShut {NoStop}%
\bibitem [{\citenamefont {Papadakis}\ \emph {et~al.}(2020)\citenamefont
  {Papadakis}, \citenamefont {Buddhiraju}, \citenamefont {Zhao}, \citenamefont
  {Zhao},\ and\ \citenamefont {Fan}}]{papadakis2020broadening}%
  \BibitemOpen
  \bibfield  {author} {\bibinfo {author} {\bibfnamefont {G.~T.}\ \bibnamefont
  {Papadakis}}, \bibinfo {author} {\bibfnamefont {S.}~\bibnamefont
  {Buddhiraju}}, \bibinfo {author} {\bibfnamefont {Z.}~\bibnamefont {Zhao}},
  \bibinfo {author} {\bibfnamefont {B.}~\bibnamefont {Zhao}}, \ and\ \bibinfo
  {author} {\bibfnamefont {S.}~\bibnamefont {Fan}},\ }\bibfield  {title}
  {\enquote {\bibinfo {title} {Broadening near-field emission for performance
  enhancement in thermophotovoltaics},}\ }\href@noop {} {\bibfield  {journal}
  {\bibinfo  {journal} {Nano Lett.}\ }\textbf {\bibinfo {volume} {20}},\
  \bibinfo {pages} {1654--1661} (\bibinfo {year} {2020})}\BibitemShut {NoStop}%
\bibitem [{\citenamefont {Messina}\ and\ \citenamefont
  {Ben-Abdallah}(2013)}]{messina2013graphene}%
  \BibitemOpen
  \bibfield  {author} {\bibinfo {author} {\bibfnamefont {R.}~\bibnamefont
  {Messina}}\ and\ \bibinfo {author} {\bibfnamefont {P.}~\bibnamefont
  {Ben-Abdallah}},\ }\bibfield  {title} {\enquote {\bibinfo {title}
  {Graphene-based photovoltaic cells for near-field thermal energy
  conversion},}\ }\href@noop {} {\bibfield  {journal} {\bibinfo  {journal}
  {Sci. Rep.}\ }\textbf {\bibinfo {volume} {3}},\ \bibinfo {pages} {1383}
  (\bibinfo {year} {2013})}\BibitemShut {NoStop}%
\bibitem [{\citenamefont {Woessner}\ \emph {et~al.}(2015)\citenamefont
  {Woessner}, \citenamefont {Lundeberg}, \citenamefont {Gao}, \citenamefont
  {Principi}, \citenamefont {Alonso-Gonz{\'a}lez}, \citenamefont {Carrega},
  \citenamefont {Watanabe}, \citenamefont {Taniguchi}, \citenamefont {Vignale},
  \citenamefont {Polini}, \citenamefont {H.}, \citenamefont {Hillenbrand},\
  and\ \citenamefont {Koppens}}]{woessner2015retime}%
  \BibitemOpen
  \bibfield  {author} {\bibinfo {author} {\bibfnamefont {A.}~\bibnamefont
  {Woessner}}, \bibinfo {author} {\bibfnamefont {M.~B.}\ \bibnamefont
  {Lundeberg}}, \bibinfo {author} {\bibfnamefont {Y.}~\bibnamefont {Gao}},
  \bibinfo {author} {\bibfnamefont {A.}~\bibnamefont {Principi}}, \bibinfo
  {author} {\bibfnamefont {P.}~\bibnamefont {Alonso-Gonz{\'a}lez}}, \bibinfo
  {author} {\bibfnamefont {M.}~\bibnamefont {Carrega}}, \bibinfo {author}
  {\bibfnamefont {K.}~\bibnamefont {Watanabe}}, \bibinfo {author}
  {\bibfnamefont {T.}~\bibnamefont {Taniguchi}}, \bibinfo {author}
  {\bibfnamefont {G.}~\bibnamefont {Vignale}}, \bibinfo {author} {\bibfnamefont
  {M.}~\bibnamefont {Polini}}, \bibinfo {author} {\bibfnamefont {James}\
  \bibnamefont {H.}}, \bibinfo {author} {\bibfnamefont {R.}~\bibnamefont
  {Hillenbrand}}, \ and\ \bibinfo {author} {\bibfnamefont {F.~H.~L.}\
  \bibnamefont {Koppens}},\ }\bibfield  {title} {\enquote {\bibinfo {title}
  {Highly confined low-loss plasmons in graphene-boron nitride
  heterostructures},}\ }\href@noop {} {\bibfield  {journal} {\bibinfo
  {journal} {Nat. Materials}\ }\textbf {\bibinfo {volume} {14}},\ \bibinfo
  {pages} {421} (\bibinfo {year} {2015})}\BibitemShut {NoStop}%
\bibitem [{\citenamefont {Zhao}\ and\ \citenamefont {Zhang}(2015)}]{Bo_JHT}%
  \BibitemOpen
  \bibfield  {author} {\bibinfo {author} {\bibfnamefont {B.}~\bibnamefont
  {Zhao}}\ and\ \bibinfo {author} {\bibfnamefont {Z.~M.}\ \bibnamefont
  {Zhang}},\ }\bibfield  {title} {\enquote {\bibinfo {title} {Enhanced photon
  tunneling by surface plasmon¨cphonon polaritons in graphene/$\rm{hBN}$
  heterostructures},}\ }\href@noop {} {\bibfield  {journal} {\bibinfo
  {journal} {J. Heat Transfer}\ }\textbf {\bibinfo {volume} {139}},\ \bibinfo
  {pages} {022701--022701--8} (\bibinfo {year} {2015})}\BibitemShut {NoStop}%
\bibitem [{\citenamefont {Zhao}\ \emph
  {et~al.}(2017{\natexlab{b}})\citenamefont {Zhao}, \citenamefont {Guizal},
  \citenamefont {Zhang}, \citenamefont {Fan},\ and\ \citenamefont
  {Antezza}}]{Bo_PRB}%
  \BibitemOpen
  \bibfield  {author} {\bibinfo {author} {\bibfnamefont {B.}~\bibnamefont
  {Zhao}}, \bibinfo {author} {\bibfnamefont {B.}~\bibnamefont {Guizal}},
  \bibinfo {author} {\bibfnamefont {Z.~M.}\ \bibnamefont {Zhang}}, \bibinfo
  {author} {\bibfnamefont {S.}~\bibnamefont {Fan}}, \ and\ \bibinfo {author}
  {\bibfnamefont {M.}~\bibnamefont {Antezza}},\ }\bibfield  {title} {\enquote
  {\bibinfo {title} {Near-field heat transfer between graphene/$\rm{hBN}$
  multilayers},}\ }\href@noop {} {\bibfield  {journal} {\bibinfo  {journal}
  {Phys. Rev. B}\ }\textbf {\bibinfo {volume} {95}},\ \bibinfo {pages} {245437}
  (\bibinfo {year} {2017}{\natexlab{b}})}\BibitemShut {NoStop}%
\bibitem [{\citenamefont {Shi}\ \emph {et~al.}(2017)\citenamefont {Shi},
  \citenamefont {Bao},\ and\ \citenamefont {He}}]{Sailing_ACS}%
  \BibitemOpen
  \bibfield  {author} {\bibinfo {author} {\bibfnamefont {K.}~\bibnamefont
  {Shi}}, \bibinfo {author} {\bibfnamefont {F.}~\bibnamefont {Bao}}, \ and\
  \bibinfo {author} {\bibfnamefont {S.}~\bibnamefont {He}},\ }\bibfield
  {title} {\enquote {\bibinfo {title} {Enhanced near-field thermal radiation
  based on multilayer graphene-hbn heterostructures},}\ }\href@noop {}
  {\bibfield  {journal} {\bibinfo  {journal} {ACS Photonics}\ }\textbf
  {\bibinfo {volume} {4}},\ \bibinfo {pages} {971--978} (\bibinfo {year}
  {2017})}\BibitemShut {NoStop}%
\bibitem [{\citenamefont {Wang}\ \emph {et~al.}(2019)\citenamefont {Wang},
  \citenamefont {Lu},\ and\ \citenamefont {Jiang}}]{PRAppliedWang}%
  \BibitemOpen
  \bibfield  {author} {\bibinfo {author} {\bibfnamefont {R.}~\bibnamefont
  {Wang}}, \bibinfo {author} {\bibfnamefont {J.}~\bibnamefont {Lu}}, \ and\
  \bibinfo {author} {\bibfnamefont {J.-H.}\ \bibnamefont {Jiang}},\ }\bibfield
  {title} {\enquote {\bibinfo {title} {Enhancing thermophotovoltaic performance
  using graphene-bn-$\mathrm{In}\mathrm{Sb}$ near-field heterostructures},}\
  }\href@noop {} {\bibfield  {journal} {\bibinfo  {journal} {Phys. Rev.
  Applied}\ }\textbf {\bibinfo {volume} {12}},\ \bibinfo {pages} {044038}
  (\bibinfo {year} {2019})}\BibitemShut {NoStop}%
\bibitem [{\citenamefont {Polder}\ and\ \citenamefont
  {Van~H.}(1971)}]{polder1971theory}%
  \BibitemOpen
  \bibfield  {author} {\bibinfo {author} {\bibfnamefont {D.}~\bibnamefont
  {Polder}}\ and\ \bibinfo {author} {\bibfnamefont {M.}~\bibnamefont
  {Van~H.}},\ }\bibfield  {title} {\enquote {\bibinfo {title} {Theory of
  radiative heat transfer between closely spaced bodies},}\ }\href@noop {}
  {\bibfield  {journal} {\bibinfo  {journal} {Phys. Rev. B}\ }\textbf {\bibinfo
  {volume} {4}},\ \bibinfo {pages} {3303} (\bibinfo {year} {1971})}\BibitemShut
  {NoStop}%
\bibitem [{\citenamefont {Pendry}(1999)}]{pendry1999radiative}%
  \BibitemOpen
  \bibfield  {author} {\bibinfo {author} {\bibfnamefont {J.~B.}\ \bibnamefont
  {Pendry}},\ }\bibfield  {title} {\enquote {\bibinfo {title} {Radiative
  exchange of heat between nanostructures},}\ }\href@noop {} {\bibfield
  {journal} {\bibinfo  {journal} {J. Phys.: Condens. Matter}\ }\textbf
  {\bibinfo {volume} {11}},\ \bibinfo {pages} {6621} (\bibinfo {year}
  {1999})}\BibitemShut {NoStop}%
\bibitem [{\citenamefont {Mulet}\ \emph {et~al.}(2002)\citenamefont {Mulet},
  \citenamefont {Joulain}, \citenamefont {Carminati},\ and\ \citenamefont
  {Greffet}}]{mulet2002enhanced}%
  \BibitemOpen
  \bibfield  {author} {\bibinfo {author} {\bibfnamefont {J.~P.}\ \bibnamefont
  {Mulet}}, \bibinfo {author} {\bibfnamefont {K.}~\bibnamefont {Joulain}},
  \bibinfo {author} {\bibfnamefont {R.}~\bibnamefont {Carminati}}, \ and\
  \bibinfo {author} {\bibfnamefont {J.~J.}\ \bibnamefont {Greffet}},\
  }\bibfield  {title} {\enquote {\bibinfo {title} {Enhanced radiative heat
  transfer at nanometric distances},}\ }\href@noop {} {\bibfield  {journal}
  {\bibinfo  {journal} {Microscale Thermophys. Eng.}\ }\textbf {\bibinfo
  {volume} {6}},\ \bibinfo {pages} {209--222} (\bibinfo {year}
  {2002})}\BibitemShut {NoStop}%
\bibitem [{\citenamefont {Whittaker}\ and\ \citenamefont
  {Culshaw}(1999)}]{scattering1999}%
  \BibitemOpen
  \bibfield  {author} {\bibinfo {author} {\bibfnamefont {D.~M.}\ \bibnamefont
  {Whittaker}}\ and\ \bibinfo {author} {\bibfnamefont {I.~S.}\ \bibnamefont
  {Culshaw}},\ }\bibfield  {title} {\enquote {\bibinfo {title}
  {Scattering-matrix treatment of patterned multilayer photonic structures},}\
  }\href@noop {} {\bibfield  {journal} {\bibinfo  {journal} {Phys. Rev. B}\
  }\textbf {\bibinfo {volume} {60}},\ \bibinfo {pages} {2610--2618} (\bibinfo
  {year} {1999})}\BibitemShut {NoStop}%
\bibitem [{\citenamefont {Zhang}(2007)}]{Zhang2007Nano}%
  \BibitemOpen
  \bibfield  {author} {\bibinfo {author} {\bibfnamefont {Z.~M.}\ \bibnamefont
  {Zhang}},\ }\href@noop {} {\emph {\bibinfo {title} {Nano/microscale heat
  transfer}}}\ (\bibinfo  {publisher} {McGraw-Hill},\ \bibinfo {year}
  {2007})\BibitemShut {NoStop}%
\bibitem [{\citenamefont {Ashcroft}\ and\ \citenamefont
  {Mermin}(1976)}]{ashcroft2010IVcurve}%
  \BibitemOpen
  \bibfield  {author} {\bibinfo {author} {\bibfnamefont {N.~W.}\ \bibnamefont
  {Ashcroft}}\ and\ \bibinfo {author} {\bibfnamefont {N.~D.}\ \bibnamefont
  {Mermin}},\ }\href@noop {} {\emph {\bibinfo {title} {Solid state physics}}}\
  (\bibinfo  {publisher} {Cengage Learning},\ \bibinfo {year}
  {1976})\BibitemShut {NoStop}%
\bibitem [{\citenamefont {Shur}(1996)}]{shur1996handbook}%
  \BibitemOpen
  \bibfield  {author} {\bibinfo {author} {\bibfnamefont {M.~S.}\ \bibnamefont
  {Shur}},\ }\href@noop {} {\emph {\bibinfo {title} {Handbook series on
  semiconductor parameters}}},\ Vol.~\bibinfo {volume} {1}\ (\bibinfo
  {publisher} {World Scientific},\ \bibinfo {year} {1996})\BibitemShut
  {NoStop}%
\bibitem [{\citenamefont {Lim}\ \emph {et~al.}(2015)\citenamefont {Lim},
  \citenamefont {Jin}, \citenamefont {Lee},\ and\ \citenamefont
  {Lee}}]{lim2015graphene}%
  \BibitemOpen
  \bibfield  {author} {\bibinfo {author} {\bibfnamefont {M.}~\bibnamefont
  {Lim}}, \bibinfo {author} {\bibfnamefont {S.}~\bibnamefont {Jin}}, \bibinfo
  {author} {\bibfnamefont {S.~S.}\ \bibnamefont {Lee}}, \ and\ \bibinfo
  {author} {\bibfnamefont {B.~J.}\ \bibnamefont {Lee}},\ }\bibfield  {title}
  {\enquote {\bibinfo {title} {Graphene-assisted si-insb thermophotovoltaic
  system for low temperature applications},}\ }\href@noop {} {\bibfield
  {journal} {\bibinfo  {journal} {Opt. Express}\ }\textbf {\bibinfo {volume}
  {23}},\ \bibinfo {pages} {A240--A253} (\bibinfo {year} {2015})}\BibitemShut
  {NoStop}%
\bibitem [{\citenamefont {Kumar}\ \emph {et~al.}(2015)\citenamefont {Kumar},
  \citenamefont {Low}, \citenamefont {Fung}, \citenamefont {Avouris},\ and\
  \citenamefont {Fang}}]{SPPPs2}%
  \BibitemOpen
  \bibfield  {author} {\bibinfo {author} {\bibfnamefont {A.}~\bibnamefont
  {Kumar}}, \bibinfo {author} {\bibfnamefont {T.}~\bibnamefont {Low}}, \bibinfo
  {author} {\bibfnamefont {K.~H.}\ \bibnamefont {Fung}}, \bibinfo {author}
  {\bibfnamefont {P.}~\bibnamefont {Avouris}}, \ and\ \bibinfo {author}
  {\bibfnamefont {N.~X.}\ \bibnamefont {Fang}},\ }\bibfield  {title} {\enquote
  {\bibinfo {title} {Tunable light-matter interaction and the role of
  hyperbolicity in graphene-$\rm{hBN}$ system},}\ }\href@noop {} {\bibfield
  {journal} {\bibinfo  {journal} {Nano Lett.}\ }\textbf {\bibinfo {volume}
  {15}},\ \bibinfo {pages} {3172--3180} (\bibinfo {year} {2015})}\BibitemShut
  {NoStop}%
\bibitem [{\citenamefont {Geick}\ \emph {et~al.}(1966)\citenamefont {Geick},
  \citenamefont {Perry},\ and\ \citenamefont {Rupprecht}}]{geick1966normal}%
  \BibitemOpen
  \bibfield  {author} {\bibinfo {author} {\bibfnamefont {R.}~\bibnamefont
  {Geick}}, \bibinfo {author} {\bibfnamefont {C.~H.}\ \bibnamefont {Perry}}, \
  and\ \bibinfo {author} {\bibfnamefont {G.}~\bibnamefont {Rupprecht}},\
  }\bibfield  {title} {\enquote {\bibinfo {title} {Normal modes in hexagonal
  boron nitride},}\ }\href@noop {} {\bibfield  {journal} {\bibinfo  {journal}
  {Phys. Rev.}\ }\textbf {\bibinfo {volume} {146}},\ \bibinfo {pages} {543}
  (\bibinfo {year} {1966})}\BibitemShut {NoStop}%
\bibitem [{\citenamefont {Caldwell}\ \emph {et~al.}(2014)\citenamefont
  {Caldwell}, \citenamefont {Kretinin}, \citenamefont {Chen}, \citenamefont
  {Giannini}, \citenamefont {Fogler}, \citenamefont {Francescato},
  \citenamefont {Ellis}, \citenamefont {Tischler}, \citenamefont {Woods},
  \citenamefont {Giles}, \citenamefont {Hong}, \citenamefont {Watanabe},
  \citenamefont {Taniguchi}, \citenamefont {Maier},\ and\ \citenamefont
  {Novoselov}}]{expara_hBN}%
  \BibitemOpen
  \bibfield  {author} {\bibinfo {author} {\bibfnamefont {J.~D.}\ \bibnamefont
  {Caldwell}}, \bibinfo {author} {\bibfnamefont {A.~V.}\ \bibnamefont
  {Kretinin}}, \bibinfo {author} {\bibfnamefont {Y.}~\bibnamefont {Chen}},
  \bibinfo {author} {\bibfnamefont {V.}~\bibnamefont {Giannini}}, \bibinfo
  {author} {\bibfnamefont {M.~M.}\ \bibnamefont {Fogler}}, \bibinfo {author}
  {\bibfnamefont {Y.}~\bibnamefont {Francescato}}, \bibinfo {author}
  {\bibfnamefont {C.~T.}\ \bibnamefont {Ellis}}, \bibinfo {author}
  {\bibfnamefont {J.~G.}\ \bibnamefont {Tischler}}, \bibinfo {author}
  {\bibfnamefont {C.~R.}\ \bibnamefont {Woods}}, \bibinfo {author}
  {\bibfnamefont {A.~J.}\ \bibnamefont {Giles}}, \bibinfo {author}
  {\bibfnamefont {M.}~\bibnamefont {Hong}}, \bibinfo {author} {\bibfnamefont
  {K.}~\bibnamefont {Watanabe}}, \bibinfo {author} {\bibfnamefont
  {T.}~\bibnamefont {Taniguchi}}, \bibinfo {author} {\bibfnamefont {S.~A.}\
  \bibnamefont {Maier}}, \ and\ \bibinfo {author} {\bibfnamefont {K.~S.}\
  \bibnamefont {Novoselov}},\ }\bibfield  {title} {\enquote {\bibinfo {title}
  {Sub-diffractional volume-confined polaritons in the natural hyperbolic
  material hexagonal boron nitride},}\ }\href@noop {} {\bibfield  {journal}
  {\bibinfo  {journal} {Nature Commun.}\ }\textbf {\bibinfo {volume} {5}},\
  \bibinfo {pages} {5221} (\bibinfo {year} {2014})}\BibitemShut {NoStop}%
\bibitem [{\citenamefont {Vakil}\ and\ \citenamefont
  {Engheta}(2011)}]{Egraphene}%
  \BibitemOpen
  \bibfield  {author} {\bibinfo {author} {\bibfnamefont {A.}~\bibnamefont
  {Vakil}}\ and\ \bibinfo {author} {\bibfnamefont {N.}~\bibnamefont
  {Engheta}},\ }\bibfield  {title} {\enquote {\bibinfo {title} {Transformation
  optics using graphene},}\ }\href@noop {} {\bibfield  {journal} {\bibinfo
  {journal} {Science}\ }\textbf {\bibinfo {volume} {332}},\ \bibinfo {pages}
  {1291--1294} (\bibinfo {year} {2011})}\BibitemShut {NoStop}%
\bibitem [{\citenamefont {Falkovsky}(2008)}]{Cgraph}%
  \BibitemOpen
  \bibfield  {author} {\bibinfo {author} {\bibfnamefont {L.~A.}\ \bibnamefont
  {Falkovsky}},\ }\bibfield  {title} {\enquote {\bibinfo {title} {Optical
  properties of graphene},}\ }\href@noop {} {\bibfield  {journal} {\bibinfo
  {journal} {J. Phys.: Conference Series}\ }\textbf {\bibinfo {volume} {129}},\
  \bibinfo {pages} {012004} (\bibinfo {year} {2008})}\BibitemShut {NoStop}%
\bibitem [{\citenamefont {Yang}\ \emph {et~al.}(2006)\citenamefont {Yang},
  \citenamefont {Cheng}, \citenamefont {Wang},\ and\ \citenamefont
  {Feng}}]{yang2006optical}%
  \BibitemOpen
  \bibfield  {author} {\bibinfo {author} {\bibfnamefont {T.-R.}\ \bibnamefont
  {Yang}}, \bibinfo {author} {\bibfnamefont {Y.}~\bibnamefont {Cheng}},
  \bibinfo {author} {\bibfnamefont {J.-B.}\ \bibnamefont {Wang}}, \ and\
  \bibinfo {author} {\bibfnamefont {Z.~C.}\ \bibnamefont {Feng}},\ }\bibfield
  {title} {\enquote {\bibinfo {title} {Optical and transport properties of insb
  thin films grown on gaas by metalorganic chemical vapor deposition},}\
  }\href@noop {} {\bibfield  {journal} {\bibinfo  {journal} {Thin solid films}\
  }\textbf {\bibinfo {volume} {498}},\ \bibinfo {pages} {158--162} (\bibinfo
  {year} {2006})}\BibitemShut {NoStop}%
\bibitem [{\citenamefont {Jacob}(2014)}]{jacob2014nanophotonics}%
  \BibitemOpen
  \bibfield  {author} {\bibinfo {author} {\bibfnamefont {Z.}~\bibnamefont
  {Jacob}},\ }\bibfield  {title} {\enquote {\bibinfo {title} {Nanophotonics:
  hyperbolic phonon-polaritons},}\ }\href@noop {} {\bibfield  {journal}
  {\bibinfo  {journal} {Nat. Mater.}\ }\textbf {\bibinfo {volume} {13}},\
  \bibinfo {pages} {1081} (\bibinfo {year} {2014})}\BibitemShut {NoStop}%
\bibitem [{\citenamefont {Brar}\ \emph {et~al.}(2014)\citenamefont {Brar},
  \citenamefont {Jang}, \citenamefont {Sherrott}, \citenamefont {Kim},
  \citenamefont {Lopez}, \citenamefont {Kim}, \citenamefont {Choi},\ and\
  \citenamefont {Atwater}}]{SPPPs1}%
  \BibitemOpen
  \bibfield  {author} {\bibinfo {author} {\bibfnamefont {V.~W.}\ \bibnamefont
  {Brar}}, \bibinfo {author} {\bibfnamefont {M.~S.}\ \bibnamefont {Jang}},
  \bibinfo {author} {\bibfnamefont {M.}~\bibnamefont {Sherrott}}, \bibinfo
  {author} {\bibfnamefont {S.}~\bibnamefont {Kim}}, \bibinfo {author}
  {\bibfnamefont {J.~J.}\ \bibnamefont {Lopez}}, \bibinfo {author}
  {\bibfnamefont {L.~B.}\ \bibnamefont {Kim}}, \bibinfo {author} {\bibfnamefont
  {M.}~\bibnamefont {Choi}}, \ and\ \bibinfo {author} {\bibfnamefont
  {H.}~\bibnamefont {Atwater}},\ }\bibfield  {title} {\enquote {\bibinfo
  {title} {Hybrid surface-phonon-plasmon polariton modes in graphene/monolayer
  h-bn heterostructures},}\ }\href@noop {} {\bibfield  {journal} {\bibinfo
  {journal} {Nano Lett.}\ }\textbf {\bibinfo {volume} {14}},\ \bibinfo {pages}
  {3876--3880} (\bibinfo {year} {2014})}\BibitemShut {NoStop}%
\end{thebibliography}%

\end{document}